\documentclass[sigconf, nonacm]{acmart}
\usepackage{amsmath}
\usepackage{mathtools}
\usepackage{graphicx} 
\usepackage{url}
\usepackage{enumitem} 

\usepackage{stmaryrd} 

\usepackage{upgreek} 
\usepackage{bbding} 

\usepackage{listings} 
\usepackage{tabularx} 
\usepackage{booktabs} 
\usepackage{multirow}
\usepackage{hhline} 
\usepackage{diagbox}
\usepackage{pgfplotstable} 

\usepackage{xcolor,colortbl}    

\usepackage{tikz} 
\usetikzlibrary{matrix}
\usetikzlibrary{calc}
\usetikzlibrary{math}
\usetikzlibrary{arrows.meta,positioning}
\usetikzlibrary{intersections, pgfplots.fillbetween}

\usepackage{easybmat} 

\usepackage{adjustbox}
\def\eox{\unskip\kern 10pt{\unitlength1pt\linethickness{.4pt}$\diamondsuit${}}} 

\usepackage{rotating}		

\usepackage{xspace} 

\usepackage{subcaption} 
\newcommand{\hide}[1]{}

\usepackage[ruled,noend,linesnumbered]{algorithm2e} 

\usepackage{setspace} 

\DontPrintSemicolon     
\SetNlSty{}{}{}                
\SetAlgoInsideSkip{smallskip}   

\SetAlFnt{\small}			
\SetAlCapFnt{\small}		
\SetAlCapNameFnt{\small}

\hypersetup{                                                            
	pdfpagemode=UseNone,               
    pdfstartview=Fit,
    pdfpagelayout=SinglePage,       
    bookmarks=false,
    bookmarksnumbered=true,
    bookmarksopen=false,             
    pdftex=true,
    unicode,
    colorlinks=true,       
    linkcolor=blue,          
    citecolor=cyan,  
    filecolor=magenta,      
     urlcolor=blue           
}

\usepackage[capitalise,nameinlink]{cleveref} 

\usepackage{aliascnt}  	

\newaliascnt{corollary}{theorem}

\aliascntresetthe{corollary}

\newaliascnt{example}{theorem}
\newtheorem{example}[example]{Example}
\aliascntresetthe{example}

\newaliascnt{definition}{theorem}

\aliascntresetthe{definition}

\newaliascnt{proposition}{theorem}

\aliascntresetthe{proposition}

\newaliascnt{lemma}{theorem}

\aliascntresetthe{lemma}

\newaliascnt{conjecture}{theorem}

\aliascntresetthe{conjecture}

\newtheorem{questionW}{Question}
\newtheorem{resultW}{Result}

{\end{itshape}
}
\AtEndPreamble{%
    \hypersetup{colorlinks,
      linkcolor=purple,
      citecolor=blue,
      urlcolor=ACMDarkBlue,
      filecolor=ACMDarkBlue}}

\usepackage{tcolorbox}		
\tcbuselibrary{breakable,skins}		

\tcbset{examplestyle/.style={
		enhanced jigsaw,	
		colback=blue!08,	
		colframe=blue!08,	
		arc=2mm,
		boxrule=0pt,		
		left=1mm,
		right=1mm,
		left skip= 0mm,  
		right skip= 0mm, 
		top=-0.5mm,		
		bottom=1mm,		
		breakable,		
		parbox = false,		
		before={\par\pagebreak[0]\vspace{1mm}\parindent=0pt},		
		after={\par\pagebreak[0]\vspace{1mm}\parindent=0pt},				
		bottomrule = 0mm,
		boxsep = 0mm,					
		topsep at break=0pt,			
		bottomsep at break=0pt,			
		pad at break=0mm,
		pad before break=1mm,		
		pad after break=1mm,		
		bottomrule at break=0mm,
		toprule at break=0mm,		
		}}

\usepackage{footnote}

\tcolorboxenvironment{example}{examplestyle}

\newcommand{\resultbox}[1]{
\begin{tcolorbox}[
	enhanced jigsaw,		
	colback=red!5,
	colframe=red!75!black,	
	arc=0mm,
	left skip=-1mm,
	right skip=-1mm,	
	left=0mm,
	topsep at break=1mm,			
	right=0mm,
	top=0mm,
	bottom=0mm,		
	breakable,		
	parbox = false		
]
\emph{#1}
\end{tcolorbox}
}

\makeatletter
\DeclareRobustCommand*\uell{\mathpalette\@uell\relax}
\newcommand*\@uell[2]{
  \setbox0=\hbox{$#1\ell$}
  \setbox1=\hbox{\rotatebox{10}{$#1\ell$}}
  \dimen0=\wd0 \advance\dimen0 by -\wd1 \divide\dimen0 by 2
  \mathord{\lower 0.1ex \hbox{\kern\dimen0\unhbox1\kern\dimen0}}
}
\makeatother

\usepackage{marginnote}

\newcommand{\introparagraph}[1]{\textbf{#1.}} 

\renewcommand{\epsilon}{\varepsilon} 

\usepackage{scalerel}
\usepackage{tikz}
\usetikzlibrary{svg.path}

\definecolor{orcidlogocol}{HTML}{A6CE39}
\tikzset{
  orcidlogo/.pic={
    \fill[orcidlogocol] svg{M256,128c0,70.7-57.3,128-128,128C57.3,256,0,198.7,0,128C0,57.3,57.3,0,128,0C198.7,0,256,57.3,256,128z};
    \fill[white] svg{M86.3,186.2H70.9V79.1h15.4v48.4V186.2z}
                 svg{M108.9,79.1h41.6c39.6,0,57,28.3,57,53.6c0,27.5-21.5,53.6-56.8,53.6h-41.8V79.1z M124.3,172.4h24.5c34.9,0,42.9-26.5,42.9-39.7c0-21.5-13.7-39.7-43.7-39.7h-23.7V172.4z}
                 svg{M88.7,56.8c0,5.5-4.5,10.1-10.1,10.1c-5.6,0-10.1-4.6-10.1-10.1c0-5.6,4.5-10.1,10.1-10.1C84.2,46.7,88.7,51.3,88.7,56.8z};
  }
}

\RequirePackage{etoolbox}
\DeclareRobustCommand\orcidicon[1]{\href{https://orcid.org/#1}{\mbox{\scalerel*{
\begin{tikzpicture}[yscale=-1,transform shape]
\pic{orcidlogo};
\end{tikzpicture}
}{|}}}}

\usepackage{array}
\usepackage{makecell}
\usepackage[utf8]{inputenc}

\newcommand{\tname}{ALT-Gen\xspace}
\newcommand{\bname}{UGEN-V1\xspace}

\newcommand{\ra}[1]{\renewcommand{\arraystretch}{#1}} %

\newcommand\vldbdoi{}
\newcommand\vldbpages{}
\newcommand\vldbvolume{}
\newcommand\vldbissue{}
\newcommand\vldbyear{2023}
\newcommand\vldbauthors{\authors}
\newcommand\vldbtitle{\shorttitle} 
\newcommand\vldbavailabilityurl{https://github.com/northeastern-datalab/alt-gen}
\newcommand\vldbpagestyle{empty} 
\begin{document}
\title{Generative Benchmark Creation for Table Union Search}


\author{Koyena Pal}
\affiliation{%
  \institution{Northeastern University}
  \city{Boston}
  \state{Massachusetts, USA}
}
\email{pal.k@northeastern.edu}

\author{Aamod Khatiwada}
\affiliation{%
  \institution{Northeastern University}
  \city{Boston}
  \state{Massachusetts, USA}
}
\email{khatiwada.a@northeastern.edu}

\author{Roee Shraga}
\affiliation{%
  \institution{Northeastern University}
  \city{Boston}
  \state{Massachusetts, USA}
}
\email{r.shraga@northeastern.edu}

\author{Ren\'ee J. Miller}
\affiliation{%
  \institution{Northeastern University}
  \city{Boston}
  \state{Massachusetts, USA}
  }
\email{miller@northeastern.edu}

\begin{abstract}
Data management has traditionally relied on synthetic data generators to generate structured benchmarks, like the TPC suite, where we can control important parameters like data size and its distribution precisely.  These benchmarks were central to the success and adoption of database management systems. But more and more, data management problems are of a semantic nature.  
An important example is finding tables that can be unioned.  While any two tables with the same cardinality can be unioned, table union search is the problem of finding tables whose union is semantically coherent.
Semantic problems cannot be benchmarked using synthetic data.  Our current methods for creating benchmarks involve the manual curation and labeling of real data.  These methods are not robust or scalable and perhaps more importantly, it is not clear how robust the created benchmarks are.
We propose to use generative AI models to 
create  structured data benchmarks for table union search.  We 
present a 
novel method for using generative models to create 
tables with specified properties.  Using this method, we create a new benchmark containing 
pairs of tables 
that are both unionable and non-unionable but related.
We thoroughly evaluate recent existing table union search methods over existing benchmarks and our new benchmark. We also present and evaluate a new table search methods based on recent large language models over all benchmarks.
We show that the new benchmark is more challenging for all methods than hand-curated benchmarks, specifically, 
the top-performing method achieves a Mean Average Precision  of around 60\%, over 30\% less than its performance on existing manually created benchmarks.  We examine why this is the case and show that the new benchmark permits more detailed analysis of methods, including a study of both false positives and false negatives that were not possible with existing benchmarks.  We discuss how our benchmark and generation method sheds more light into the successes and failures of table union search methods  sparking new insights into further research on more robust approaches.
\end{abstract}
\maketitle

 \pagestyle{\vldbpagestyle}
\begingroup\small\noindent\raggedright\textbf{PVLDB Reference Format:}\\
\vldbauthors. \vldbtitle. PVLDB, \vldbvolume(\vldbissue): \vldbpages, \vldbyear.\\
\href{https://doi.org/\vldbdoi}{doi:\vldbdoi}
\endgroup
\begingroup
\renewcommand\thefootnote{}\footnote{\noindent
This work is licensed under the Creative Commons BY-NC-ND 4.0 International License. Visit \url{https://creativecommons.org/licenses/by-nc-nd/4.0/} to view a copy of this license. For any use beyond those covered by this license, obtain permission by emailing \href{mailto:info@vldb.org}{info@vldb.org}. Copyright is held by the owner/author(s). Publication rights licensed to the VLDB Endowment. \\
\raggedright Proceedings of the VLDB Endowment, Vol. \vldbvolume, No. \vldbissue\ %
ISSN 2150-8097. \\
\href{https://doi.org/\vldbdoi}{doi:\vldbdoi} \\
}\addtocounter{footnote}{-1}\endgroup

\ifdefempty{\vldbavailabilityurl}{}{
\vspace{.3cm}
\begingroup\small\noindent\raggedright\textbf{PVLDB Artifact Availability:}\\
The source code, data, and/or other artifacts have been made available at \url{\vldbavailabilityurl}.
\endgroup
}

\section{Introduction}

David Paterson states that “when a field has good benchmarks, we settle debates and the field makes rapid progress”~\cite{patterson2012technical}.  
Traditionally, benchmark generation is done using synthetic data generators that precisely control parameters such as data size, data distributions, and correlations in data~\cite{DBLP:books/mk/Gray93}.
These parameters influence the standard DBMS performance metrics: response time and through-put. However today, more and more data management challenges require not only the fast and scalable processing of data, but also the understanding of semantics. A common example that we 
consider in this 
paper is \emph{table union} -- do two tables contain attributes and relationships with the same semantics so that they can be meaningfully unioned~\cite{nargesian2018table,2023_khatiwada_santos}. For this problem, the most important performance metrics are accuracy and robustness over different data types.

As the semantics of structured data is important in evaluation, synthetic data generators are ineffective. As a result, researchers find and curate (label) real datasets manually. 
Nargesian et al. created  labeled 
benchmarks (which we will call TUS-Small and TUS-Large in this work) using real open data (from government open data portals)~\cite{nargesian2018table} which has been reused 
for diverse applications, not just the table union problem~\cite{DBLP:conf/icde/BogatuFP020, DBLP:conf/icde/KoutrasSIPBFLBK21, DBLP:conf/edbt/LeventidisRMRG21}. 
They took several large tables and sliced them horizontally and vertically to create tables that are unionable on some or all attributes. This approach was later extended by Khatiwada et al.~\cite{2023_khatiwada_santos} (SANTOS-Small and SANTOS-LARGE) and used in other work~\cite{DBLP:journals/pvldb/FanWLZM23, 2023_helali_linked_data, 2023_hu_autotus}.
Another example is the T2K Gold Benchmark, which was created from a large web table corpus~\cite{WDC_t2d} matched with properties from DBpedia and hand labeled. All these efforts require considerable human effort to find appropriate real data and to label the ground truth. For example, while TUS-Large benchmark~\cite{nargesian2018table} offers thousands of ``labeled'' tables, they originate from only 10 seed tables which 
are necessarily limited in the variety of semantics they capture (for example, the topics and data types in the tables). The SANTOS-Large~\cite{2023_khatiwada_santos}, which is large in size and based on real open data tables, offers only 80 human-labeled samples. Also, none of these benchmarks offer a set of non-unionable table pairs and assumes that table pairs that are not labeled as unionable are not unionable.  This assumption is made because the original tables are very different.
Despite these limitations, the hand-labeled benchmarks are still valuable to the community and have been reused widely.

Importantly, it is not possible to systematically vary many important independent parameters (such as varying the type or semantics of data to test robustness, the problem complexity, or data size) using these human-gathered and hand-labeled benchmarks. In addition, the need for new benchmarks also comes from the effectiveness of state-of-the-art models (Starmie~\cite{DBLP:journals/pvldb/FanWLZM23}) which stands at around 95\% MAP over existing benchmarks. It is possible that these benchmarks are too easy.

Generative AI models are machine learning algorithms that are tasked to predict the next text or related image (or other types of output) based on their input instruction. 
They have become increasingly popular in recent years, especially in NLP, where models such as GPT3~\cite{brown2020language} and ChatGPT~\cite{openaichat}
are used by millions of users on a daily basis. These models can be used as is (aka zero-shot learning~\cite{xian2018zero}) or by providing a (small) set of examples that steer the model in the right direction with respect to the task at hand (aka In-Context Learning (ICL)~\cite{brown2020language,dong2022survey}).
Based on their performance on such tasks, these models could provide the key to making innovative new advancements in benchmarking structured data.

\noindent\textbf{Contributions:} In this paper, we employ the capabilities of generative AI models to create and improve benchmark datasets for the table union task. Our contributions can be summarized as follows.
\begin{enumerate}
    \item \textbf{Automated LLM Table Generator (\tname):} an automated framework to generate tables for table union search tasks based on a set of desired properties.
    \item \textbf{A New Table Union Search Benchmark (\bname):} we generate and share a new table union search benchmark featuring 1050 tables. The benchmark offers a variety of 50 topics ranging from World Geography to Veterinary Medicine, 1000 labeled table pairs (500 unionable/500 non-unionable) allowing fine-grained effectiveness analysis.
    \item \textbf{Table Union Search Evaluation:}
    we evaluate and analyze existing table union search methods over our new benchmark and the existing TUS and SANTOS benchmarks.   While the best search methods achieve a MAP of over 90\% on existing benchmarks, we show that the new benchmark is  more challenging.  The best MAP values are around 60\%, potentially triggering research into more robust  methods.  
    \item \textbf{A More Thorough Experimental Analysis:} previous work has used hand-labeled benchmarks as-is.  We show that our \tname methodology lends itself to more robust benchmarking by considering other important parameters.  Specifically, we vary the sparsity of the \bname benchmark (how many null values it contains) and evaluate methods on a variety of topics.
    We show that by using \tname, we can do ablation studies on these parameters to understand the table union search methods even better, something that has not been done before.  
    \item \textbf{New Table Union Search Method:}
    in generating a benchmark using an LLM for table union search, an important question is whether an LLM is better at table union search (especially on its own benchmark) than the best existing methods.   
    An LLM can easily be used to classify two tables as unionable or non-unionable, but not for table union search (given a query table, find tables within a massive repository that are unionable with it).  So to answer this question, we present a new table union search method that uses a state-of-the-art search technique to find a set of candidates tables (Starmie~\cite{DBLP:journals/pvldb/FanWLZM23}) and then uses an LLM to classify table pairs among the candidates as unionable/non-unionable. 
    Our in depth analysis of the performance of this new method with existing methods highlights some important and interesting areas for future work and improvement.
\end{enumerate}

In Section~\ref{sec:related}, we survey related work. Section~\ref{sec:benchmarkgen} presents our methodology to generate benchmarks and the new benchmark. Section~\ref{sec:exp} provides the analysis and evaluation.

\section{Related Work}\label{sec:related}

We first present the state-of-the-art methods in table union search (including methods that appear in our experimental study).  We then consider related work on benchmark creation in data management tasks, including benchmark generation for semantic tasks like data cleaning and data integration. Finally, we introduce generative models that we use for our new benchmark creation methodology and the way they are used in the literature. 

\subsection{Table Union Search} 
Given a query table by the user, 
table union search techniques intend to find the data lake tables that can add new rows to the query table~\cite{2023_khatiwada_santos}. We cover several search techniques that were developed over the last few years aiming at finding 
unionable tables over the data lakes. 
\citet{nargesian2018table}  considered two tables to be unionable if they have a subset of unionable columns. The column unionability is determined by using an ensemble of three statistical tests based on value overlap, semantic overlap, and word embedding of the column values.
\textbf{D$^3$L}~\cite{DBLP:conf/icde/BogatuFP020} extended that work~\cite{nargesian2018table} by considering, in addition to value overlap and word embedding measures, three additional attribute unionability measures (column header similarity, numerical value distribution, and regular expression similarity). \textbf{SANTOS}~\cite{2023_khatiwada_santos} considered the similarity of both columns and the binary relationship between the columns to make table unionability decisions. The binary relationships help SANTOS to understand table context better and omit the unioning of the tables having similar columns but different contexts~\cite{2023_khatiwada_santos}. 
\textbf{Starmie}~\cite{DBLP:journals/pvldb/FanWLZM23} uses a contrastive-learning approach to capture the contexts of the entire table rather than just the binary relationships, and use them to search for the unionable tables. Starmie captures the table context in the form of column embeddings. In very recent work, \citet{2023_hu_autotus} used the contextualized representation of the relationship between the column pairs to capture the table contexts and use them to find the unionable tables from the data lakes (however, open code is not available for this work).


The benchmarks used (and created) in the respective papers and their limitations are discussed in the introduction. Importantly, their creation and evaluation involve a significant amount of manual annotation. In this work, we present a principled way to use generative models, aiming to eliminate such manual workload during benchmark creation.






\subsection{Benchmark Generation in Databases}

Benchmark generation has been considered for many data management tasks.
In data cleaning, \citet{2015_arocena_bart} introduced BART which can be used to add errors into clean databases and evaluate data-repairing algorithms.
In data integration, iBench~\cite{2015_arocena_ibench} was introduced to generate schemas and schema constraints with arbitrarily large and complex mappings.
These systems start with real data and metadata and systematically vary specific parameters that can influence the performance of cleaning or integration systems.  In a similar spirit, we show how \tname can generate realistic data and how we can use \tname to vary certain parameters (such as textuality) during the generation process.  We also consider how to post-process the generated result to vary other parameters such as sparsity.  

Others have used knowledge graphs~\cite{2021_cutrona_semtab}, Git repositories~\cite{2023_hulsebos_gittables}, and the web~\cite{brinkmann2023web} to generate tabular benchmarks.  
For example, the SemTab challenge~\cite{2021_cutrona_semtab}, which focuses on the semantic annotation of the tabular datasets using knowledge graphs, generates benchmarks using Wiki data
for important intra-table tasks such as column type annotation and column-column relationship annotation. However, limited value coverage in the knowledge graph~\cite{2023_khatiwada_santos} could pose a challenge for creating diverse benchmarks and this is an area where generative models could be helpful. Another example is The Web Data Commons project~\cite{brinkmann2023web} which extracts schema.org~\cite{guha2016schema} data from the Web using Common Crawl~\cite{commoncrawl}. This project yielded several successful benchmark for multiple tasks such as Product Matching. The extraction is limited to a specific schema and still requires manual annotation.  

A recent data discovery benchmark includes Lakebench which contains 8 open datasets for joinability and unionability tasks~\cite{srinivas2023lakebench}. Different from our work where we generate tables for \emph{table search} task using a generative model, Lakebench benchmarks existing open datasets and new datasets created using Knowledge graphs, for fine tuning large language models.

\subsection{Generative Models and LLMs}
Generative models are machine learning models that are trained to learn their training data distribution. Based on this learned pattern and given input, they generate text, images, videos, and more. They have been used to augment training data in various related tasks such as commonsense reasoning~\cite{yang2020generative}, event detection~\cite{Veyseh2021AugmentingOE}, text classification and summarization~\cite{ chintagunta2021medically}. More recently, researchers have coined the term large language models (LLMs) which refers to generative language models in general. In this paper, we use Open AI's GPT3~\cite{brown2020language} for benchmark generation and also experiment with GPT2-xl, Alpaca~\cite{alpaca} and Vicuna~\cite{vicuna} which are open source via Hugging Face~\cite{wolf2020transformers}.

LLMs have been used within the data management community extensively. For example,~\citet{arora2023language} use LLMs to generate structured views of semi-structure data lakes. Others use LLMs to extract knowledge graphs~\cite{west-etal-2022-symbolic, cohen-etal-2023-crawling}. Trummer used GPT-3 to generate code for query processing~\cite{trummer2022bert,trummer2022codexdb}. Recently there were some additional initial attempts to solve other data management tasks such as information extraction~\cite{brinkmann2023product} and entity matching~\cite{peeters2023using} using prompting and in-context learning. But to the best of our knowledge, generative models have not be used in generating benchmarks for semantic data management tasks.

\section{Generative Union Benchmarks}\label{sec:benchmarkgen}
\begin{figure*}
    \centering
\includegraphics[scale = 0.55]{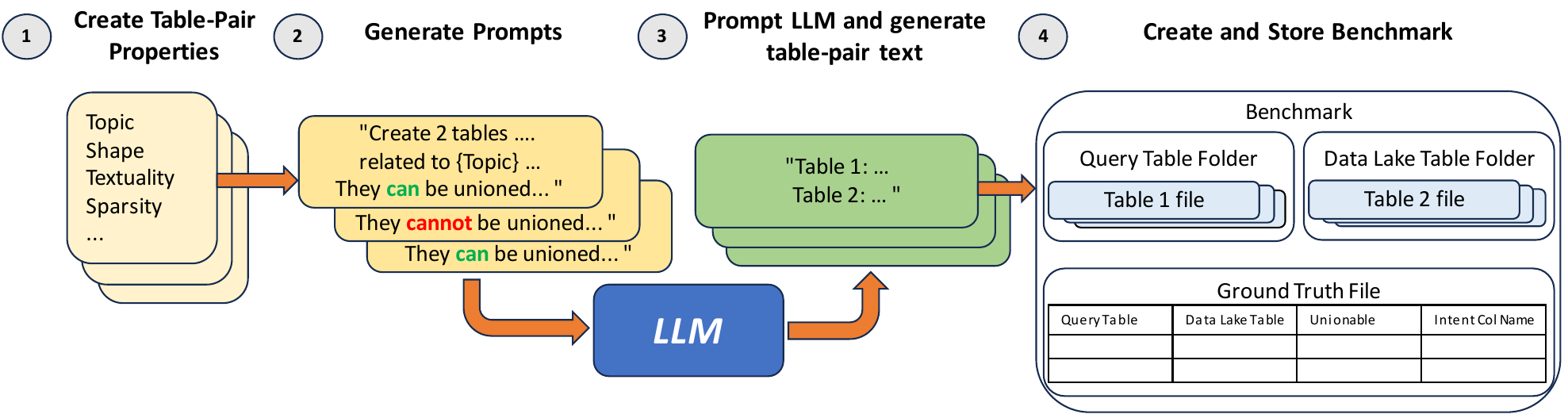}
    \caption{Pipeline for generating benchmarks through Large Language Models}
    \label{fig:pipeline}
\end{figure*}
Before presenting our benchmark generation method, we first consider what properties a good benchmark for table union search should have.  We then present our generative benchmark methodology \tname and 
our new benchmark \bname. 





\subsection{Desired Properties for Union Benchmarks}\label{sec:properties}


Following Primpeli and Bizer~\cite{primpeli2020profiling}, who presented a set of properties that should be considered within (hand curated) entity matching benchmarks, we present a set of properties the union benchmarks should have.  The first (realism) is a required property.  Beyond this, for the other properties, we argue that a benchmark generator should be able to vary the degree or value of the property systematically.

\noindent$\bullet$ \textbf{Realism:} the ability to generate ``realistic'' and semantically meaningful (not necessarily ``real'') tuples and structured tables.  Due to this property, all union benchmarks to date have been created from real open data sets rather than using synthetic data.  We require that a benchmark generator be able to create realistic data rather than randomly associating values as a synthetic data generator might.  \\
\noindent$\bullet$ \textbf{Typing:} the ability to generate a variety of types of values (e.g., numeric and textual).  This is important as solutions for semantic data management problems often differ based on types.  A benchmark generator should be able to generate tables with a subset of types (for example, only categorical values), or with varying degrees of different types (for example, half numeric and half strings).  \\
\noindent$\bullet$ \textbf{Incompleteness:} the ability to generate incomplete data containing nulls.  A benchmark generator should be able to create complete tables and systematically vary the degree of incompleteness (number of nulls) contained in a table.

To evaluate whether the benchmark generator is able to achieve these properties, we generate a benchmark through this generator and evaluate the effectiveness of the benchmark itself. We adopt the ~\citet{primpeli2020profiling} dataset profiling approach (for ours and existing benchmarks) and consider the following factors:

\noindent$\bullet$ \textbf{Schema Complexity:} This factor refers to the number of attributes contributing to solving the unionability task. \\
\noindent$\bullet$ \textbf{Textuality:} This refers to the average length of words of the 
attribute values. \\
\noindent$\bullet$ \textbf{Sparsity:} This dimension refers to the ratio of missing attribute values in the tables considered for unionability. \\ 
\noindent$\bullet$ \textbf{Development Set Size:} This refers to the amount of positive and negative record pairs in the
benchmark for the unionability task.\\
\noindent$\bullet$ \textbf{Corner Cases:} This factor refers to the ratio of false positive and false negative cases, which are referred to as corner cases. This is a method-specific that largely looks at the ratio of challenging cases in a benchmark for each method tested on the benchmark. \\
We propose to use a generative model (specifically an LLM) to automate annotation and ensure inter-table unionability, thereby improving benchmark diversity and scalability.
The benchmark will be comprised of two sets of tables, representing data lake and query tables. A ground truth file will also be included, which will provide information about which data lake tables can be unioned with each given query table and which cannot.

\subsection{Unionability Benchmark Generation}\label{sec:benchmarkgentemplate}
Existing hand-curated benchmarks are fairly homogeneous and relatively small (the largest fully labeled benchmark has 5,000 tables~\cite{nargesian2018table}) as they are generated from a few large \emph{seed} tables that are then sliced (horizontally and vertically) to create unionable tables~\cite{nargesian2018table}. Using generative models, we do not have to rely on \emph{seed} tables. Instead, we will navigate the model to generate pairs of unionable (and non-unionable) tables via prompting and in-context learning. 

\subsubsection{\tname: Automated LLM Table Generator}\label{sec:benchmarkgentemplate}
\Cref{fig:pipeline} illustrates the pipeline of generating tables using LLMs.
We prompt an LLM (for instance, GPT3) to generate pairs of tables with the following set of features customized for each table: topic, number of rows and columns (shape), and 
the textuality rate. To generate tables for a unionability benchmark, we add additional properties to the prompt such as whether the table pairs are supposed to be unionable or not. In cases where they are unionable, we prompt the LLM to generate both the tables as well as the key column name in the query table (table 1), which act as the primary column containing shared data points with a paired data lake (table 2) column.\footnote{A key or intent column is required by one of the search techniques that we will evaluate~\cite{2023_khatiwada_santos}.}  
\resultbox{
    \texttt{Create 2 tables with cells separated by |. Table 1 has \{$T1_{row}$\} rows, \{$T1_{col}$\} columns and \{${T1_{textuality}}$\} words, related to \{${T1_{topic}}$\}. Table 2 has \{$T2_{row}$\} rows, \{$T2_{col}$\} columns and \{${T2_{textuality}}$\} words, related to \{${T2_{topic}}$\}.}
}
\noindent In the prompt \{$T1_{row}$\}, \{$T1_{col}$\}, \{${T1_{textuality}}$\} and \{${T1_{topic}}$\} (and similarly for $T2$) are parameters. These parameters are currently instantiated randomly during the `create table-pair properties' phase in \Cref{fig:pipeline}, but can be set by users if a table-pair needs to have certain user-specific conditions. 
Depending on whether these tables should be unioned or not, we add either of the relevant sentences to the prompt.
\resultbox{
\texttt{ They can be unioned because they have only \{$T_{unionable\_col}$\} semantically common columns, and at least 1 related row values across the tables.}
}
\resultbox{
\texttt{  They cannot be unioned because they have $0$ semantically common columns.}
}
The unionability-specific instructions can be switched to another inter-table task (such as entity matching or joinability). For instance, in the case of entity matching, we could state that the data points ``can be entity-matched" or ``cannot be entity-matched." This generalizes the pipeline while creating specialized tables for various inter-table tasks.

To allow for increased consistency in the format of the output generated by an LLM
, we add the following instruction to the prompt template:
\resultbox{
    \texttt{Answer the above task in the following format: \\
            Table 1: \{table 1\} \\
            Table 2: \{table 2\} \\
            Key: \{key column in Table 1\}} 
}
\noindent In this prompt, ``Key:..." is included only if the table pair should be unionable. Note that relevant methods can use this key when necessary, for example, SANTOS~\cite{2023_khatiwada_santos} can use the key as the intent column of the table.

After we create this prompt, we input this to the LLM, which then outputs the tables (and key column value, if unionable) as a string. We can then implement a post-processing script to convert the string-formatted tables into CSV files. 

Apart from the size of the tables and the benchmark itself, we also consider the sparsity present in each table. Because this feature involves removing information instead of adding any new information, we automate the process of generating such tables through an algorithm that replaces cell values with a null (empty) value. Although sparse tables can be generated by LLM, it is more consistent and cost-efficient to create non-sparse tables using the model and multiple sparse ones using an algorithm. 

\begin{figure*}[h]
    {
   \begin{minipage}[t]{0.5\textwidth}
    \includegraphics[width=\linewidth]{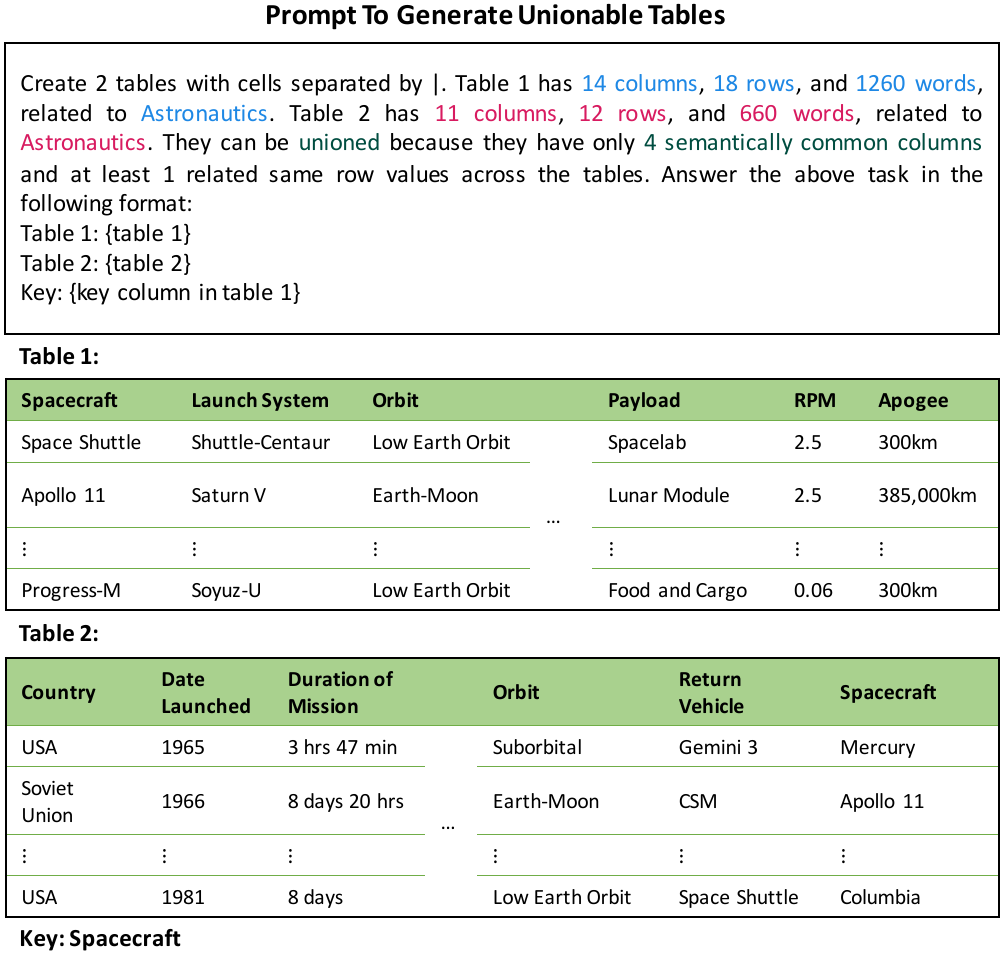}
    \end{minipage}%
    \begin{minipage}[t]{0.5\textwidth}
    \includegraphics[width=\linewidth]{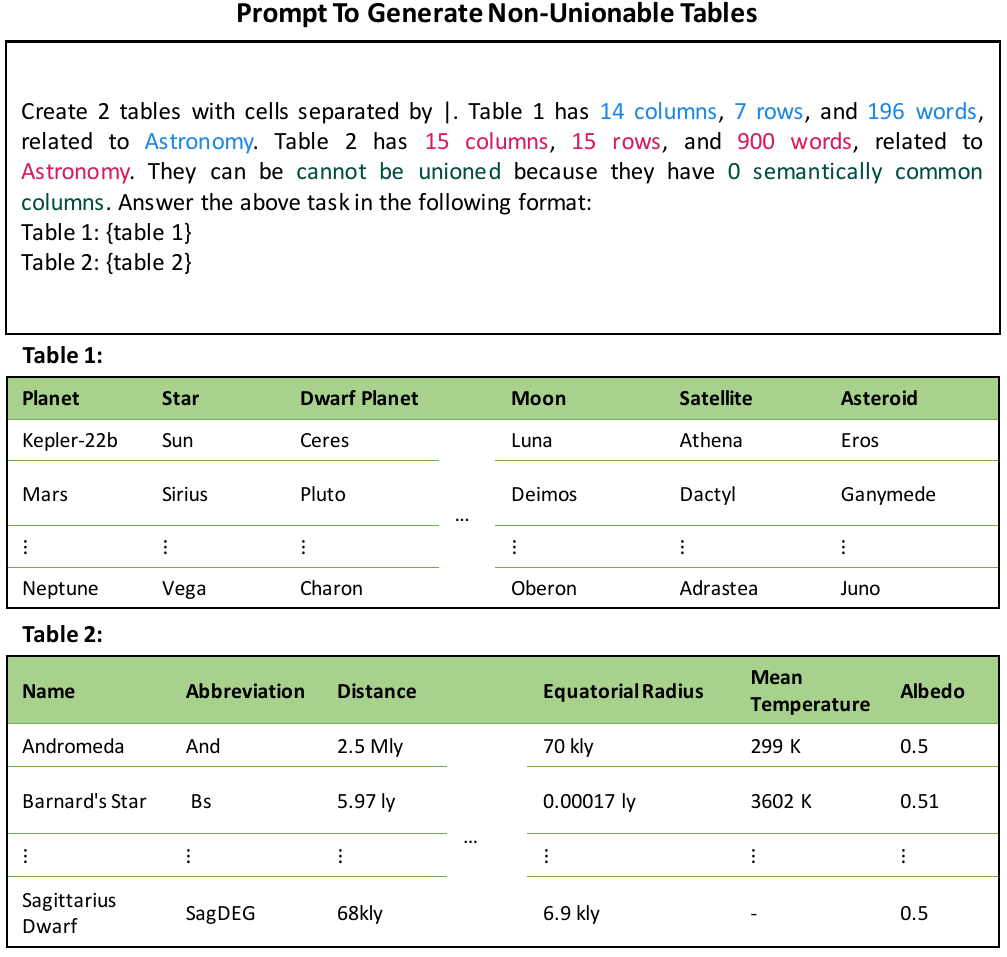}
    \end{minipage}%
    \caption{Example of generating pairs of unionable and non-unionable table pairs through \tname.}
    \label{fig:generation-example}
    }
\end{figure*}

To create diverse and realistic benchmark, we prompt the LLM to generate 
tables of different topics. An important property, which is not 
present
in other benchmarks, is the ability to generate non-unionable tables of the same topic. 


Unlike hand-curated benchmarks where any pair of tables  that are not explicitly created to be unionable (that is two tables that are created from slicing and dicing the same large seed table) are assumed to be non-unionable, our generation method, \tname, 
explicitly creates pairs of non-unionable tables.  
If each seed table is selected to be very different from the others (and on a different topic) as done in both the TUS~\cite{nargesian2018table} and SANTOS~\cite{2023_khatiwada_santos} benchmarks that we compare against, then the non-unionable tables may be unrealistically easy to determine.
The benefits of our approach is we get varied examples from a variety of topics that would allow for more realistic unionable and non-unionable table-pair examples (including tables on the same topic as a query table that do not union with it) that we can test for unionability.   This creates opportunities for current and future table union search method to consider harder cases. 
We now provide an example illustrating prompts and the corresponding generated tables.

\begin{example}
\label{ex:running_example}
    \Cref{fig:generation-example} provides two generation examples including a prompt and a pair of tables. On the left a pair of unionable tables is generated and on the right a pair of non-unionable tables is generated. Each prompt provides a set of properties for the model to satisfy and whether the generated tables can (left) or cannot (right) be unioned. In this case both tables contain real data. For example, \texttt{Apollo 11}'s launch system is \texttt{Saturn V} and its orbit is \texttt{Earth-Moon}. Potentially the generative model can generate realistic data that is not real. For example, a record about the \texttt{Hubble Space} spacecraft with the launch system \texttt{Solaris Rocket} that has a \texttt{Low Earth Orbit} and \texttt{Orion MPCV} payload. While the orbit and payload are real, the Hubble Space spacecraft and its launch system are made up. This can be important in real-world applications where real data can be considered as a privacy leakage~\cite{carlini2021extracting,carlini2022quantifying}.
    

    The unionable tables (left) are related to \texttt{Astronautics} and share (at least) 4 unionable (semantically common) columns, out of which \texttt{Spacecraft} and \texttt{Orbit} are visible in the figure. The provided key of the tables (denoted under Table 2) is \texttt{Spacecraft}. 
    Notably, the columns overlap (e.g., \texttt{Appollo 11}) but not completely and contain multiple types (e.g., \texttt{RPM} is numeric and \texttt{Date Launched} is a date). 

    The non-unionable tables (right) are both related to \texttt{Astronomy} and do not share any unionable (semantically common) columns. Importantly, these two tables are not unionable but are of the same topic i.e. \texttt{Astronomy}. 
    Such samples are not labeled explicitly in the traditional benchmarks.
    We envision such examples to be the most challenging ones for contemporary methods which are built on top of identifying semantics, which can be mistakenly thought of as topic. Also here we see several data types such as distances (\texttt{Distance} and \texttt{Equatorial Radius}) and, interestingly missing values (last row in \texttt{Mean Temperature}). The latter being another property we aimed for in a benchmark (incompleteness). 
\end{example}

\subsubsection{\bname: A New Unionability Benchmark}\label{sec:new_unionability_benchmark}
\begin{table*}
 \ra{1.1}
\caption{Overview of Union Table Search Benchmarks}
\begin{tabular}{@{}ll l l l l l@{}}
\toprule
\multicolumn{2}{@{}l@{}}{\textbf{Benchmark}} &
  \textbf{SANTOS-Small} &
  \textbf{SANTOS-Large} &
  \textbf{TUS-Small} &
  \textbf{TUS-Large} &
  \textbf{UGEN-V1} \\ 
  \hline
\multicolumn{2}{@{}l@{}}{\textbf{\# Tables}}                                                     & 550             & 11166      & 1655 & 4295 & 1050          \\ 

\hline
\multicolumn{1}{@{}l@{}}{\multirow{2}{*}{\makecell[l]{\textbf{\# Average Shape}\\ \texttt{rows x columns}}}} &
  \textbf{Query} &
  21401.70 x 12.30 &
  12917.23 x 12.71 &
  4423.20 x 12.88 &
  1996.22 x 11.85 &
  8.34 x 11.00 \\ 
\multicolumn{1}{@{}l@{}}{}                                          & \textbf{Datalake}             & 6921.38 x 11.49 & 7684.54 x 10.98      & 4457.10 x 9.68 & 1996.22 x 11.85 & 7.81 x 10.41\\ 
\hline
\multicolumn{1}{@{}l@{}}{\multirow{2}{*}{\textbf{Size}}}            & \textbf{Query}                & 127.7 MB        & 143.1 MB & 139 M & 1.5 GB & 205KB       \\ 
\multicolumn{1}{@{}l@{}}{}                                          & \textbf{Datalake}             & 442.4 MB        & 11.48 MB & 1.2 GB & 1.5 GB & 4.1MB       \\ 
\hline
\multicolumn{1}{@{}l@{}}{\multirow{2}{*}{\textbf{\# Labeled Pair}}} & \textbf{Unionable}            & 695             & 80       & 230921 & 1202826 & 500          \\ 
\multicolumn{1}{@{}l@{}}{}                                          & \textbf{Non-unionable}        & 0               & 0        & 0 & 0 & 500          \\ 
\hline
\multicolumn{1}{@{}l@{}}{\multirow{8}{*}{\makecell[l]{\textbf{Attributes}\\ \texttt{(query,datalake)}}}} &
  \textbf{\# Short Str} &
  (79, 568) &
  (60, 8541) &
  (168,3752) &
  15740 &
  (5, 193) \\ 
\multicolumn{1}{@{}l@{}}{}                                          & \textbf{\# Medium Str}        & (90, 1048)      & (198, 26081)      & (156,1785) & 10394 & (56, 904)    \\ 
\multicolumn{1}{@{}l@{}}{}                                          & \textbf{\# Long Str}          & (299, 2995)     & (491, 55313)      & (1107,7909) & 21092 & (441, 8559)  \\ 
\multicolumn{1}{@{}l@{}}{}                                          & \textbf{\# Num}               & (139, 1496)     & (169, 22350)      & (179,1364) & 3208 & (32, 400)    \\ 
\multicolumn{1}{@{}l@{}}{}                                          & \textbf{Avg. Density} & (99.94\%, 99.92\%)  & (99.83\%, 99.78\%)      & (99.94\%,99.96\%) & 99.87\% & (94.02\%, 93.50\%)          \\ 
\multicolumn{1}{@{}l@{}}{}                                          & \textbf{\# Sparsely Unique}   & (481, 4777) & (764, 89375)      & (1173,10505) & 39787 & (388, 6463)          \\ 
\multicolumn{1}{@{}l@{}}{}                                          & \textbf{\# Moderately Unique} & (117, 1380) & (240, 28938)      & (428,3623) & 9177 & (161, 3943)          \\ 
\multicolumn{1}{@{}l@{}}{}                                          & \textbf{\# Densely Unique}    & (17, 165)   & (13,3371)      & (9,682) & 1928 & (1, 3)          \\ 
\bottomrule
\end{tabular}%

\label{tab:union-overview}
\end{table*}

Using GPT3 (as the LLM model) and prompt template described in \Cref{sec:benchmarkgentemplate}, we generated 1000 pairs of tables (half unionable, half non-unionable), which took approximately 10 hours and \$18 USD. These tables covered 50 topics ranging from World Geography and Art History to Genealogy and Veterinary Medicine. For each topic, we generated 1 query table and 20 other data lake tables that are either unionable or non-unionable with the query table. If we create $I$ versions of each table with different levels of sparsity then we have $1050 * I$ tables and at least $1000 *  I$ labeled pairs of tables in the benchmark.  For example, if $I = 5$, we create 5 versions of each table which may, for illustration, include a complete table, one with 5\% nulls, one with 10\% nulls, one with 15\% nulls, and one with 20\% nulls.

\bname and our generation code \tname are publicly available at \url{https://github.com/northeastern-datalab/alt-gen}. 



\subsection{Comparison of Unionability Benchmarks}
In ~\cref{tab:union-overview}, we provide an overview of the new benchmark (\bname) as well as other unionability benchmarks. We provide  the following features for each benchmark:
\begin{enumerate}
    \item \# Tables: Total number of tables in the benchmark.
    \item \# Average Shape: Average number of rows and columns in the tables, including query tables and data lake tables.
    \item Size: The total size/storage space of the benchmark.
    \item \# Labeled Pairs: For each benchmark, there is a ground truth file that indicates which pairs of tables are unionable or not. It can be a non-exhaustive list since non-unionability, for instance, can be inferred by permuting and pairing all other tables that are not stated to be unionable (though this may include false negatives). Nevertheless, we record the number of unionable and non-unionable pairs mentioned in the ground truth file.
    \item Attributes: In the attributes section of the table, we give the following properties of the query tables (first number) and data lake tables (second numbe) in the benchmarks:
    \begin{enumerate}
        \item \# Short Str: Number of columns in the benchmark that have an average length of strings that is less than 3.
        \item \# Medium Str: Number of columns in the benchmark that have an average length of strings less than 6, but greater than or equal to 3.
        \item \# Long Str: Number of columns in the benchmark that have an average length of strings greater than or equal to 6.
        \item \# Num: Number of columns in the benchmark with a numerical data type.
        \item Avg Density: Average percentage of non-null values present in columns of the benchmark. 
        \item \# Sparsely Unique: Number of columns that have less than 10\% of the unique values in the benchmark.
        \item \# Moderately Unique: Number of columns that have less than 50\% but greater than 10\% of the unique values in the benchmark.
        \item \# Densely Unique: Number of columns that have greater than 50\%, of the unique values present in the benchmark.
    \end{enumerate}
\end{enumerate}

One of the unique features included in the new benchmark is the presence of labeled non-unionable pairs. Past benchmarks have inferred non-unionability by considering pairs of tables that are not otherwise labeled as unionable. While we can infer additional non-unionable tables in the \bname benchmark as well, we have labeled non-unionable pairs that allow us to perform more in-depth analysis with respect to non-unionability (see \Cref{sec:nonunionability-analysis}). Compared to past benchmarks, \bname has the smallest size of tables in terms of average number of rows and columns in both query and data lake sets of tables. Despite its size, it is a more challenging benchmark as shown in the next section. Unlike previous benchmarks, \bname has higher textuality rate, which is evident in the comparatively higher long string ratio, and higher sparsity, which is indicated with a comparatively lower density value. These are contributing factors that make \bname more difficult. Finally, we note that the new benchmark also provides topics and, specifically, non-unionable table pairs of the same topic.  


\section{Analysis and Evaluation}\label{sec:exp}
We now perform empirical analysis over the existing benchmarks and our new benchmark created using generative model. We start by explaining the evaluation tasks~(\cref{sec:tasks}) followed by the details on experimental setups~(\cref{section:experimental_setup}), competing methods~(\cref{section:baselines}), benchmarks used for evaluation~(\cref{section:exp_benchmark}) and evaluation measures~(\cref{section:evaluation_measures}). Then we report the effectiveness of different methods in table union search task~(\cref{section:union_search_effectiveness}) followed by different ablation studies~(\cref{section:ablation_study}).

\subsection{Tasks}\label{sec:tasks}


We analyze the efficacy of the benchmark for the following two tasks.   
The first is the traditional search problem for which there are numerous solutions in the literature~\cite{nargesian2018table,DBLP:conf/icde/BogatuFP020,2023_khatiwada_santos,DBLP:journals/pvldb/FanWLZM23}.

\introparagraph{Table Union Search:}~\cite{nargesian2018table,2023_khatiwada_santos} Given a query table, $Q$ and a set of data lake tables, $\mathcal T = \{t_1, t_2, ...t_m\}$ where $|\mathcal T| = m$, find the top-$k$ data lake tables in $\mathcal T$ that are unionable with $Q$.


\introparagraph{Union Table Classification}

Given a pair of tables $T_1$ and $T_2$, determine if they are unionable or not. The second problem is motivated by our use of LLMs to generate a unionability benchmark.  Since we are claiming LLMs can produce unionable tables, can we also use an LLM to classify whether two tables we give to it are unionable or not.  For example, we can use labeled pair of tables from our own \bname benchmark or hand-labeled pairs from any of the existing benchmarks. The latter can be viewed as a sanity check on whether the notion of unionability used by an LLM is a reasonable one conforming to definitions used in current research.

\subsection{Experimental Setup}
\label{section:experimental_setup}
We run experiments using Python 3.8 on a server with A100 80GB PCIe. We host the large language models locally in the aforementioned GPU. For the models unavailable to host locally, we use their open APIs. 
Our experiments aim to answer specific questions.

\begin{itemize}
    \item How do the current union search methods perform on the new \bname benchmark created using the generative model?
    \item Can generative models understand unionability?
    \item How is the performance of union search techniques impacted by the sparsity (see ~\cref{sec:properties})?
    \item How is the performance of union search techniques on specific topics?
\end{itemize}

\subsection{Union Search Methods} 
\label{section:baselines}
We evaluate the publicly available recent union search methods over the existing and new benchmarks.

\introparagraph{D$^3$L~\cite{DBLP:conf/icde/BogatuFP020}} Bogatu et al. extended TUS~\cite{nargesian2018table} to use not only word embeddings, knowledge graph mappings, or value overlap, but also
column header similarity, distributions for numerical columns, and regular expressions.
We implement $D^3L$ using publicly available code.\footnote{\url{https://github.com/alex-bogatu/d3l}} For a fair comparison, we do not use the column header similarity metric since the existing benchmarks and other baselines are schema-agnostic.

\introparagraph{SANTOS~\cite{2023_khatiwada_santos}} SANTOS uses column semantics and the semantics of relationships between column pairs to search for the unionable tables. 
SANTOS only uses column values and does not use metadata like column headers. To find column semantics and relationships semantics, SANTOS uses an external knowledge base and a synthesized knowledge base created using the data lake itself. To run SANTOS, we use the public code provided with the paper.\footnote{\url{https://github.com/northeastern-datalab/santos}}

\introparagraph{Starmie~\cite{DBLP:journals/pvldb/FanWLZM23}} Starmie is a recent state-of-the-art self-supervised table union search technique based on contrastive learning. Starmie captures the table contexts in the form of column embeddings and uses them to perform table union search. We reproduced Starmie following the instruction in its open implementation.\footnote{\url{https://github.com/megagonlabs/starmie}}


\introparagraph{Starmie-LLM} 
The LLMs cannot be used directly for the search task but can as mentioned be used for union classification.  Therefore, to assess LLMs in the table union search task, we use an existing table union search method to search for a set of candidate unionable tables for each query table. Then we prompt LLMs to classify whether the query table with each candidate table are unionable or not. This two-phase approach is very common for information retrieval applications~\cite{dehghani2017neural,shraga2020web} in which two models are applied consecutively. In our experimental setup, we use 
Starmie to search for a reasonable number of candidate unionable tables. Then we prompt an LLM to classify each query-candidate table pairs are truly unionable by asking the following question:
\resultbox{
    \texttt{Are the following tables unionable? Answer in the following format: \\
            Unionable: \{yes/no\}}
}
We use recent LLMs that have shown promising performance in other generative tasks. 
Specifically, we use GPT2-xl\footnote{\url{https://huggingface.co/gpt2-xl}}, GPT3 (text-davinci-003 model with 175 billion parameters)\footnote{ \url{https://platform.openai.com/docs/models/gpt-3}}, Alpaca (7 Billion parameters)\footnote{\url{https://huggingface.co/circulus/alpaca-7b}}, and Vicuna (7 billion parameters)\footnote{\url{https://huggingface.co/lmsys/vicuna-7b-v1.3}} models in our experiments.
We denote respective LLM variations using Starmie-GPT2-XL, Starmie-GPT3, Starmie-Alpaca, and Starmie-Vicuna.

LLMs can function either as they are (zero-shot) or can be directed towards specific tasks by providing a few examples (in-context learning). For comparison with other methods, we utilize Starmie-LLM's zero-shot versions denoted by Starmie-LLM$_{\text{Zero}}$ and the in-context versions with an optimal number of examples (Starmie-LLM$_{\text{Optim}}$), which gives the best performance against its other in-context versions.


\subsection{Benchmarks}
\label{section:exp_benchmark}
Along with the newly created \textbf{\bname} benchmark, we use the existing union search benchmarks for our experiments.
The details of each benchmark are given in~\cref{tab:union-overview}.
We use \textbf{SANTOS-Small} and \textbf{TUS-Small} benchmarks to measure the effectiveness of union search methods. Note that we do not use \textbf{SANTOS-Large} because its ground truth is not annotated~\cite{2023_khatiwada_santos}. 
We also exclude the 
\textbf{TUS-Large} from the experiments as its tables do not have intent columns annotated, a requirement for the SANTOS method.

\subsection{Evaluation measures}
\label{section:evaluation_measures}
Following the previous table union search work~\cite{nargesian2018table, 2023_khatiwada_santos, 2023_hu_autotus}, we use Precision@k (P@k), Recall@k (R@k) and Mean Average Precision (MAP@k) to evaluate the effectiveness of table union search techniques. We use the binary label for unionability (unionable/non-unionable), and each benchmark's ground truth contains a set of unionable data lake tables for each query table.

Given a Query Table $Q$, let $\mathcal{T}_{Q}$ represent the ground truth set of its unionable tables and $\hat{\mathcal{T}}_{Q}$ represents the set of top-k unionable tables returned by some method. Then, $P@k$, $R@k$ and $MAP@k$ with respect to $Q$ are given by~\cite{2023_khatiwada_santos}:
\begin{equation}\label{eq:P_R}
    P@k = \frac{\mathcal{T}_{Q}\cap\hat{\mathcal{T}}_{Q}}{\hat{\mathcal{T}}_{Q}}, 
    R@k = \frac{\mathcal{T}_{Q}\cap\hat{\mathcal{T}}_{Q}}{\mathcal{T}_{Q}}, 
     MAP@k = \frac{1}{|\hat{\mathcal{T}}_{Q}|}\sum_{k=1}^{|\hat{\mathcal{T}}_{Q}|} P@k
\end{equation}
Consistent with the prior work~\cite{2023_khatiwada_santos, DBLP:journals/pvldb/FanWLZM23}, we run experiments with k less than the ground truth size to ensure that there are enough true results available in the data lake when searching for top-k unionable tables per query table.  In such cases, a Recall@k of 1 is not possible since all unionable tables cannot be returned.

Furthermore, unlike existing benchmarks, the new \bname Benchmark also contains the labeled non-unionable pairs. Therefore, we also measure Accuracy (ACC) and Corner Case Ratio (CCR) in this benchmark. 
Let, TP, FP, TN, and FN represent the True Positives, False Positives, True Negatives, and False Negatives by a method over a benchmark. Then,
\begin{equation}\label{eq:accuracy}
    ACC = \frac{TP + TN}{TP + TN + FP + FN},
    CCR = 1 - ACC
\end{equation}


 



\noindent We further use a confusion matrix to illustrate the specific TP, FP, TN, and FN values. 

\subsection{Union Search Effectiveness}
\label{section:union_search_effectiveness}
\begin{table}
\caption{$\text{P@k}$, $\text{MAP@k}$ and $\text{R@k}$ of table union search methods over different benchmarks.}
\begin{tabular}{@{}lllll@{}}
\toprule
\multicolumn{1}{@{}l@{}}{\textbf{Benchmark}} &
  \textbf{Method} &
  \textbf{$MAP@k$} &
  \textbf{$P@k$}  & 
  \textbf{$R@k$} \\ 
  \hline
  \multicolumn{1}{@{}l@{}}{\multirow{5}{*}{\makecell[l]{\textbf{TUS-Small}\\ \texttt{k=60}}}} &
  D$^3$L &
  0.79 &
  0.77 &
  0.21 \\ 
  \multicolumn{1}{@{}l@{}}{} &
  SANTOS &
  0.88 &
  0.81 &
  0.23 \\ 
  \multicolumn{1}{@{}l@{}}{} &
  Starmie &
  \textbf{0.94} &
  \textbf{0.82} &
  \textbf{0.27} \\
   \multicolumn{1}{@{}l@{}}{} &
  Starmie-Vicuna$_{\text{Zero}}$ &
    0.89 &
    0.70 &
    0.24 \\
  \multicolumn{1}{@{}l@{}}{} &
  Starmie-Vicuna$_{\text{Optim}}$ &
    \textbf{0.94} &
    0.80 &
    \textbf{0.27} \\
  \hline
  \multicolumn{1}{@{}l@{}}{\multirow{5}{*}{\makecell[l]{\textbf{SANTOS-Small}\\ \texttt{k=10}}}} &
  D$^3$L &
  0.52 &
  0.58 &
  0.42 \\ 
  \multicolumn{1}{@{}l@{}}{} &
  SANTOS &
  0.94 &
  \textbf{0.91} &
  \textbf{0.69} \\
  \multicolumn{1}{@{}l@{}}{} &
  Starmie &
  \textbf{0.95} &
  \textbf{0.91} &
  0.68 \\
  \multicolumn{1}{@{}l@{}}{} &
  Starmie-Vicuna$_{\text{Zero}}$ &
   0.81 &
   0.68 &
   0.50 \\
  \multicolumn{1}{@{}l@{}}{} &
  Starmie-Vicuna$_{\text{Optim}}$ &
   \textbf{0.95} &
   0.90  &
   0.67 \\
    \hline
  \multicolumn{1}{@{}l@{}}{\multirow{5}{*}{\makecell[l]{\textbf{UGEN-V1}\\ \texttt{k=10}}}} &
  D$^3$L &
  0.26 &
  0.19 &
  0.19 \\ 
  \multicolumn{1}{@{}l@{}}{} &
  SANTOS &
  0.56 &
  0.46 &
  0.46 \\ 
  \multicolumn{1}{@{}l@{}}{} &
  Starmie &
  \textbf{0.61} &
  \textbf{0.51} &
  \textbf{0.51} \\
   \multicolumn{1}{@{}l@{}}{} &
  Starmie-Vicuna$_{\text{Zero}}$ &
    0.44 &
    0.32 &
    0.32 \\
  \multicolumn{1}{@{}l@{}}{} &
  Starmie-Vicuna$_{\text{Optim}}$ &
    0.57 &
    0.48 &
    0.48 \\
\bottomrule
\end{tabular}%
\label{tab:result-overview}
\end{table}

\begin{figure}[h]
    {
    \centering
    \begin{minipage}[t]{\textwidth}
\includegraphics[width=0.48\linewidth]{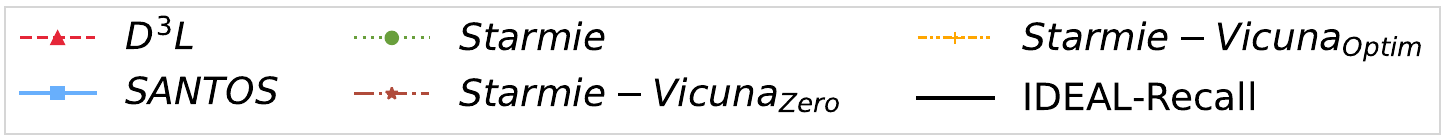}
    \end{minipage}%
    }
    \subfloat[Average $P@k$ on TUS-Small]{
    \begin{minipage}[t]{0.49\linewidth}
    \includegraphics[width=\linewidth]{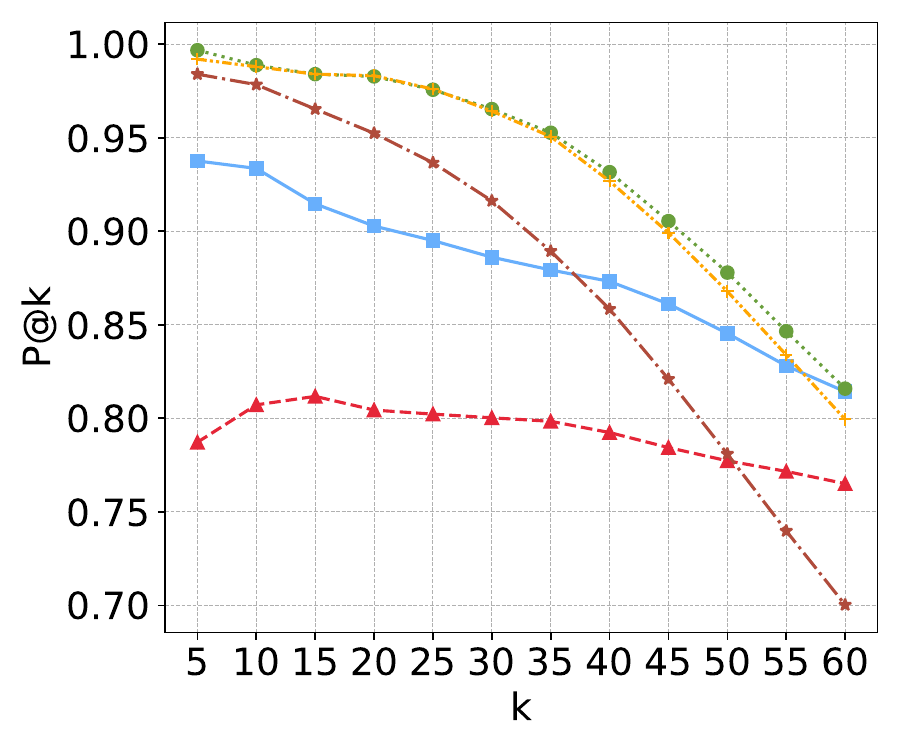}
    \end{minipage}%
    }
    \subfloat[Average $R@k$ on TUS-Small]{
    \begin{minipage}[t]{0.49\linewidth}
    \includegraphics[width=\linewidth]{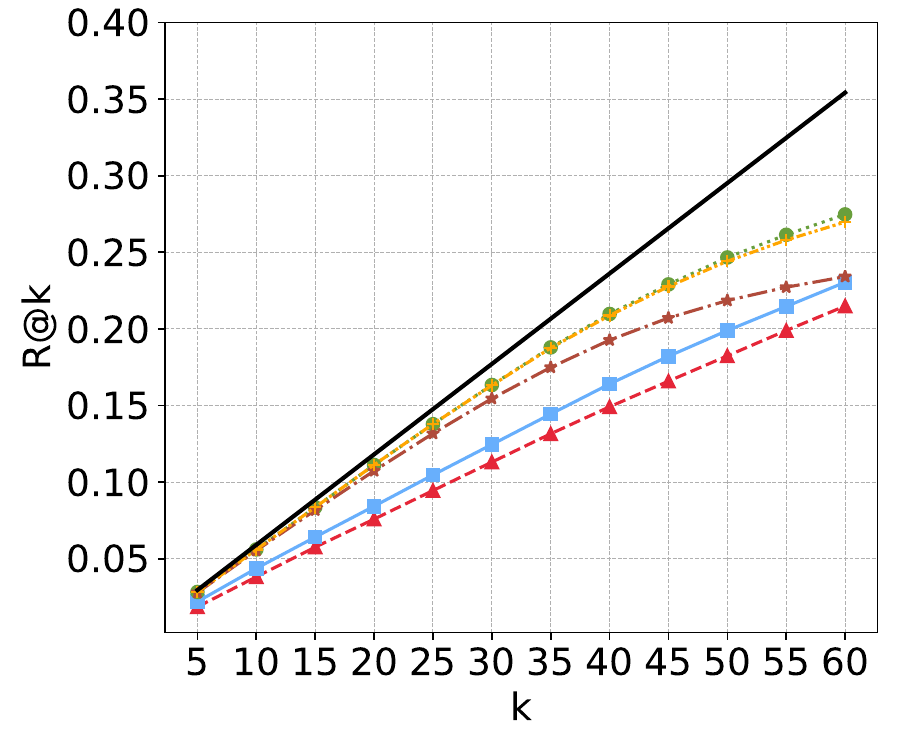}
    \end{minipage}%
    }
    \hfill
    \subfloat[Average $P@k$ on Santos-Small]{
   \begin{minipage}[t]{0.49\linewidth}
    \includegraphics[width=\linewidth]{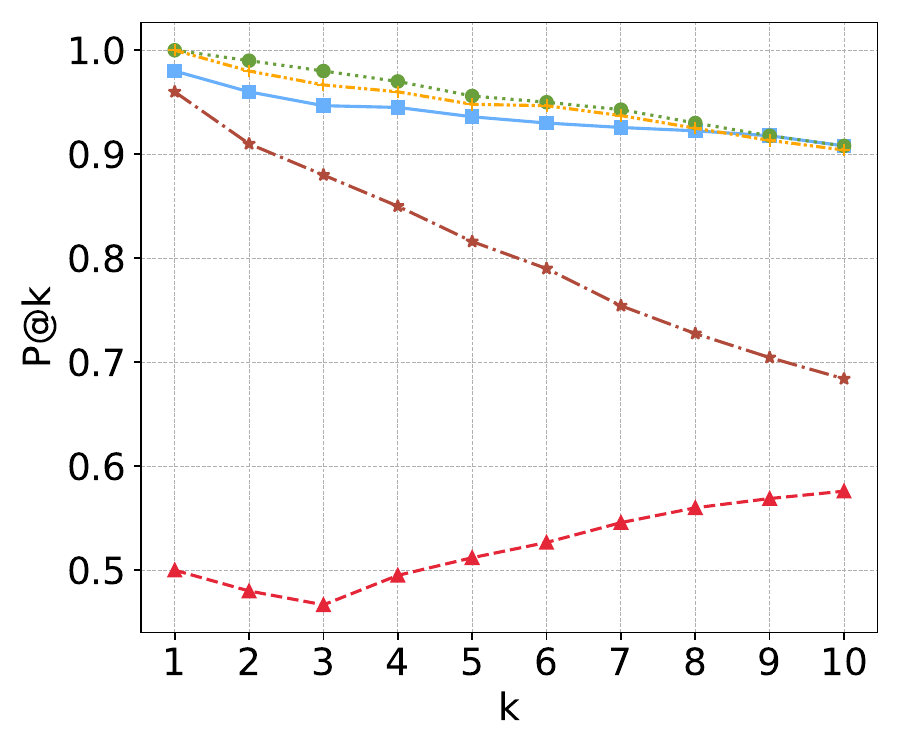}
    \end{minipage}%
    }
    \subfloat[Average $R@k$ on Santos-Small]{
    \begin{minipage}[t]{0.49\linewidth}
    \includegraphics[width=\linewidth]{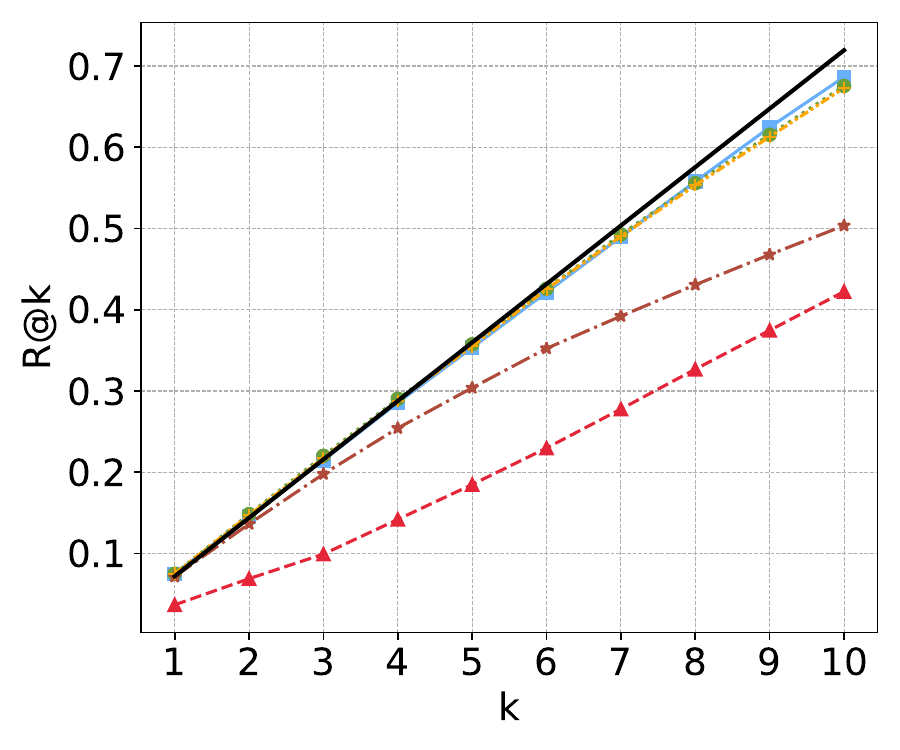}
    \end{minipage}%
    }
    \hfill
    \subfloat[Average $P@k$ on \bname]{
   \begin{minipage}[t]{0.49\linewidth}
    \includegraphics[width=\linewidth]{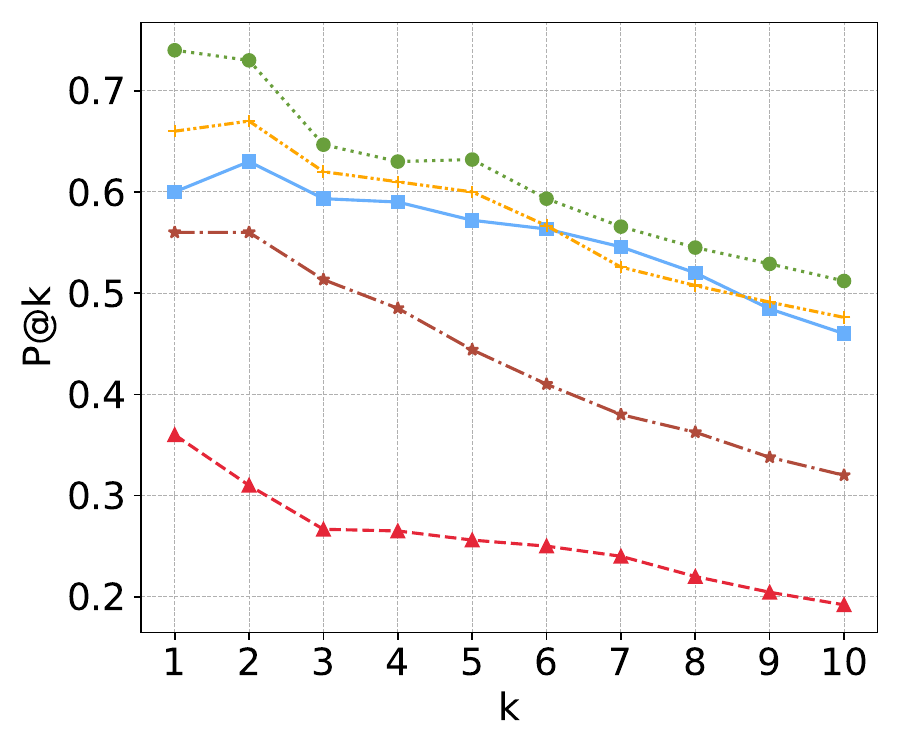}
    \end{minipage}%
    }
    \subfloat[Average $R@k$ on \bname]{
    \begin{minipage}[t]{0.49\linewidth}
    \includegraphics[width=\linewidth]{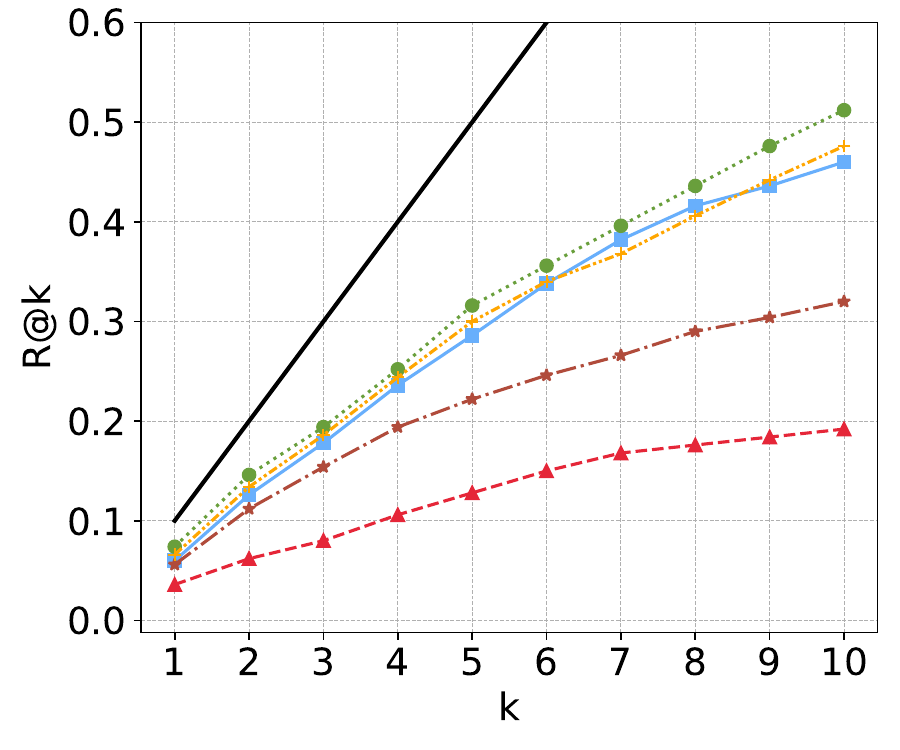}
    \end{minipage}%
    }
    \caption{Effectiveness of baselines in different benchmarks}
    \label{fig:pr_rc_all}
\end{figure}

Now we evaluate the performance of various table union search methods (described in \cref{section:baselines}) using both existing and new benchmarks~(detailed in \cref{section:exp_benchmark}). 
Specifically, we 
compare D$^3$L, SANTOS, Starmie, and Starmie-LLMs. 
Among Starmie-LLMs, we select the best-performing Starmie-Vicuna's zero-shot version (Starmie-Vicuna$_{\text{Zero}}$) and 
optimal-shot version (Starmie-Vicuna$_{\text{Optim}}$) as it shows a balanced performance in both versions. 
In the ablation study (\cref{section:ablation_study}), we 
discuss the results of other Starmie-LLM variations and their different versions.
The evaluation metrics used are MAP@k, P@k, and R@k.
Following previous work~\cite{2023_khatiwada_santos, nargesian2018table}, the maximum value of k is chosen for each benchmark, based on the number of unionable tables available in the data lake for query tables. For TUS-small, we select k up to 60. For SANTOS-small, we go up to k = 10. 
The effectiveness result for the maximum value of k on each benchmark is presented in \cref{tab:result-overview}. We \textbf{bold} the score of the best-performing method on each measure and benchmark.
Further results for other smaller values of k are plotted in \cref{fig:pr_rc_all} with different values of k in horizontal axes, the evaluation metrics in vertical axes, and the methods encoded using different colors and line styles. As noted, the recall cannot be perfect if k is smaller than the ground truth size~\cite{2023_khatiwada_santos}. So, we show the IDEAL-RECALL line that indicates the maximum possible recall for each value of k.



\introparagraph{Existing Benchmarks} The performances of all methods follow a similar trend in both TUS-Small and SANTOS-small benchmarks. Specifically, Starmie mostly stands out in terms of MAP@k, P@k, and R@k followed by 
Starmie-Vicuna$_{\text{Optim}}$, Starmie-Vicuna$_{\text{Zero}}$,
SANTOS and D$^3$L. 
 As Starmie captures the entire table context using contrastive learning, it seems to understand the table semantics better in both benchmarks.
 Furthermore, when we steer Starmie-Vicuna to understand unionability by providing an optimal number of in-context examples, its MAP increases by 5 \% over the zero-shot version in the TUS-small benchmark, and by 14 \% in the SANTOS-small benchmark matching the performance of the best-performing Starmie. This indicates that the out-of-the-box LLMs are not good enough to understand unionability, but they can be taught to do so by using in-context learning.
 Moreover, SANTOS also shows a comparable performance against Starmie and Starmie-Vicuna$_{\text{Optim}}$ in both benchmarks and is the best method in SANTOS-small in terms of P@10 and R@10. Mainly, SANTOS seems to be benefited from the use of the binary relationships between the columns which helps it to better understand the table context. 

\introparagraph{\bname} In the new \bname benchmark, Starmie continues to outperform all other methods across all three evaluation metrics, followed by Starmie-Vicuna$_{\text{Optim}}$, SANTOS, Starmie-Vicuna$_{\text{Zero}}$, and D$^3$L. However, it is important to note that the performance of all methods shows a significant drop when compared to their performance on the SANTOS-Small and TUS-Small benchmarks.
For instance, Starmie's MAP@k on 
\bname for all values of k drops by over 33 \% compared to its MAP@k on both the SANTOS-Small and TUS-Small benchmarks. This drop in performance can be attributed to the difficulty the methods face in distinguishing non-unionable tables from tables within the same topic.  Note the other benchmarks 
do not include this case due to the way they slice and dice tables each of which is from a different topic (see \cref{fig:generation-example}).
In contrast, \bname contains both unionable and non-unionable table pairs for each query table.
This scenario is quite common in data lakes, but it is challenging to include in hand labeled benchmarks~\cite{2023_khatiwada_santos, nargesian2018table}. 
Hence, the systematic benchmark creation using a generative model enables more control over the diversity of table topics and facilitates the construction of a challenging yet realistic benchmark.
In \cref{section:ablation_study}, we conduct an error analysis and delve into the performance of different techniques in each domain.

Finally, the lower performance achieved by Starmie-Vicuna, the best among the Starmie-LLM variations, in comparison to the classical table union search methods, indicates that even advanced 
LLMs 
like Vicuna 
face challenges in the new 
benchmark. This shows ample opportunity for 
creating better table union search methods.


\subsection{Ablation Study}
\label{section:ablation_study}

Now we perform fine-grained analyses over the new benchmark to get better insights into the impact of different factors in the table unionability search.

\subsubsection{In-Context learning}

\begin{figure}[t]
    {
    \centering
    \begin{minipage}[t]{\textwidth}
    \hspace{1.5cm}
    \includegraphics[width=0.3\linewidth]{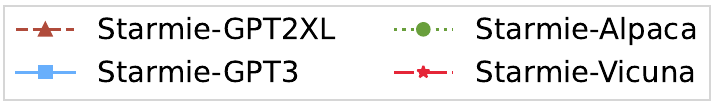}
    \end{minipage}%
    }
    \subfloat[Average $MAP@k$ on \bname]{
    \begin{minipage}[t]{0.49\linewidth}
    \includegraphics[width=\linewidth]{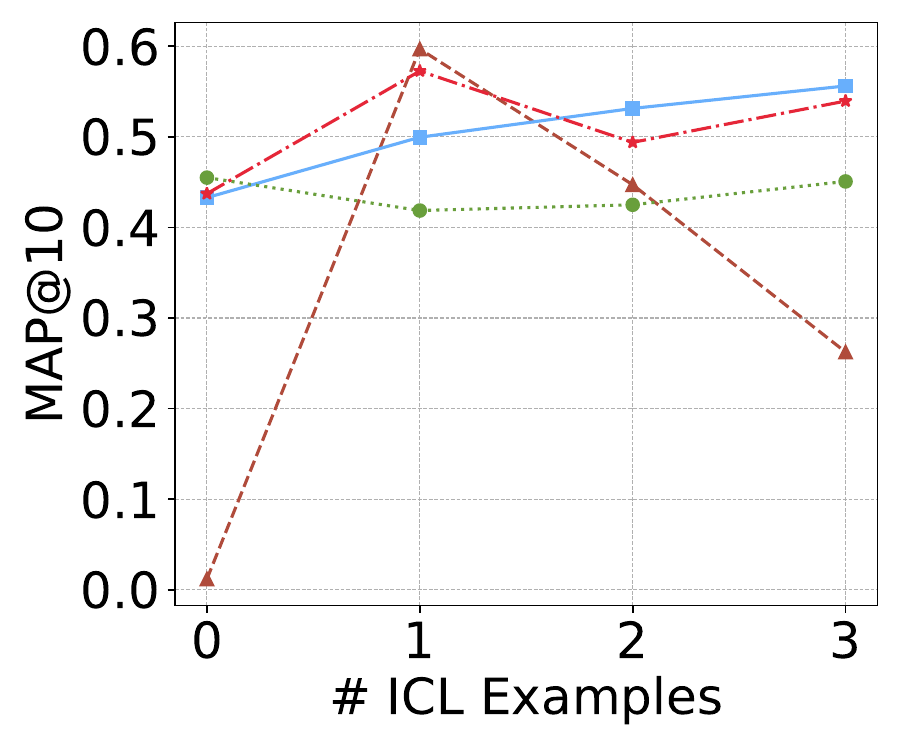}
    \end{minipage}%
    }
    \subfloat[Average $P@k$ on \bname]{
    \begin{minipage}[t]{0.49\linewidth}
    \includegraphics[width=\linewidth]{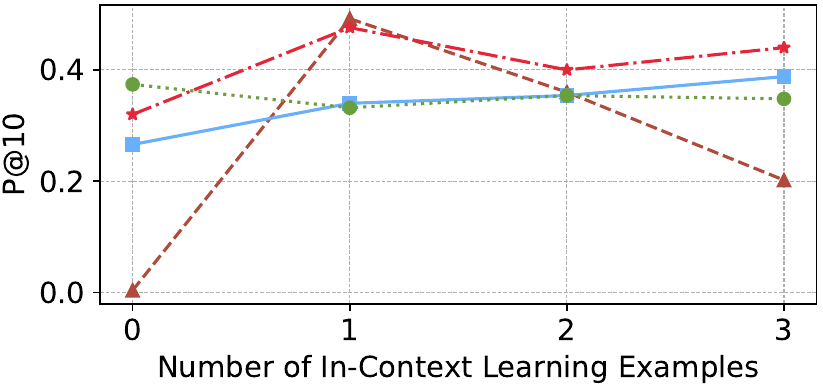}
    \end{minipage}%
    }
    \caption{$\text{MAP@10}$ and $\text{P@10}$ for different numbers of in-context learning examples provided during Starmie-LLM prompting for identifying unionability in \bname.}
    \label{fig:icl-overview}
\end{figure}

In-Context learning is 
used in LLMs where a model is able to learn a task based on a few demonstrations provided in the prompt~\cite{brown2020language}. This method does not require any model parameter training or updates, which makes it a quick adoption strategy to improve a model's performance. However, this method is sensitive to the number, order, and content of the examples~\cite{zhao2021calibrate}. Hence, we follow a variation of the setup described by~\citet{liu2021makes} to create good in-context learning examples. This variation uses the Roberta-Large model~\cite{liu2019roberta}\footnote{\url{https://huggingface.co/roberta-large}} 
to encode the potential in-context examples, and employs average Euclidean distance for each example with respect to all query examples of the \bname benchmark. We sort these distance values and get $n$ closest distance values to create $n$ examples for in-context learning. 

To prevent data leakage, we conduct evaluations on both the \bname and TUS-Small benchmarks, employing in-context examples exclusively from the SANTOS-Small benchmark. Likewise, when testing the methods on the SANTOS-Small benchmark, we exclusively utilize examples from the TUS-Small benchmark. Each example in this evaluation consists of a pair of either unionable or non-unionable tables.

In \Cref{fig:icl-overview}, we illustrate how in-context learning affects the effectiveness of the Starmie-LLM method on the \bname Benchmark. On the horizontal axis, we vary the number of examples given to the method and in the vertical axis, we show the performance metrics  MAP@10 and P@10.
Note that P@10 and R@10 are the same because we report performance at k = 10, which is the number of unionable tables in ground truth for each query table, hence we only show P@10.
We encode each method using different colors and line styles.
As the performance trend of each method is almost the same in all the metrics, we focus our discussion on MAP@10 (top-most graph). 

All but Starmie-Alpaca benefit significantly when they are provided with an in-context example to understand unionability. For instance, Starmie-GPT2-XL's MAP@10 increased by 59 \% when provided with an in-context example (one-shot learning). This implies that GPT2-XL possibly does not understand table unionability, but learns very easily with examples. Similarly, the one-shot version of Starmie-GPT3 and Starmie-Vicuna also enjoy an increase of up to 13 \% in MAP@10 with one-shot learning. Notice however, the performance of all the models saturates or even goes down if we provide more examples. For example, Starmie-GPT2-XL's MAP@10 drops from 60 \% on one-shot learning to 26 \% on three-shot learning. 
This indicates that the LLM models can learn unionability with examples but may start to underfit the provided examples if we increase their numbers.

\subsubsection{Sparsity}
\label{section:sparsity}

\begin{figure}[t]
    {
    \centering
    \begin{minipage}[t]{\textwidth}
\includegraphics[width=0.48\linewidth]{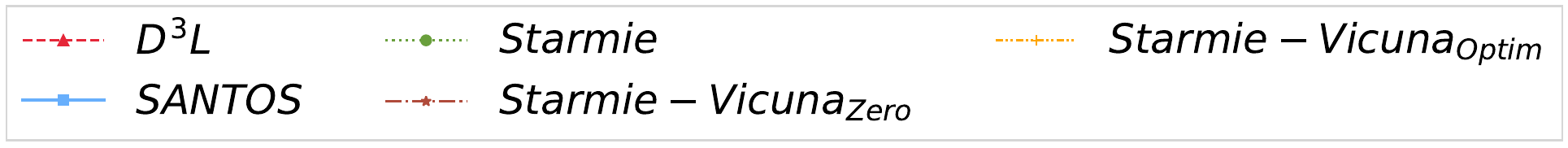}
    \end{minipage}%
    }
    \subfloat[Average $MAP@k$ on \bname]{
    \begin{minipage}[t]{0.49\linewidth}
    \includegraphics[width=\linewidth]{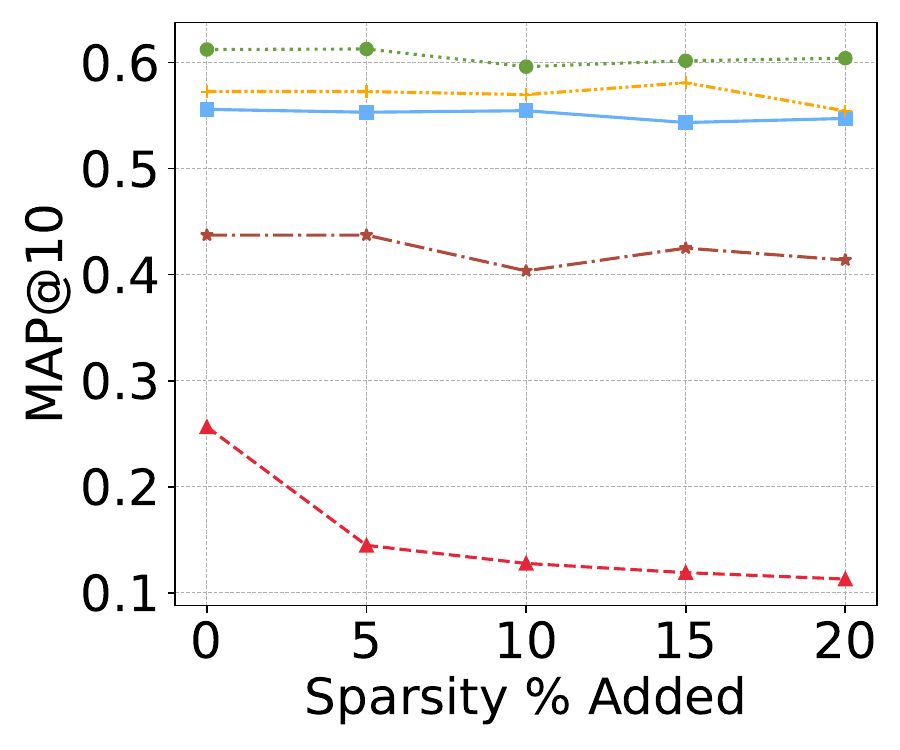}
    \end{minipage}%
    }
    \subfloat[Average $P@k$ on \bname]{
    \begin{minipage}[t]{0.49\linewidth}
    \includegraphics[width=\linewidth]{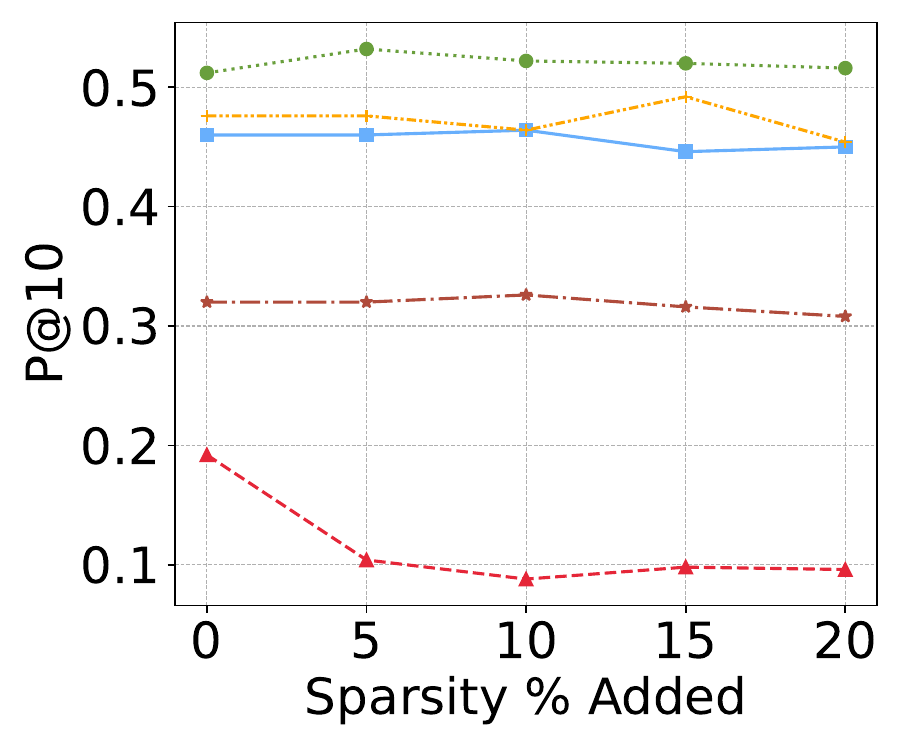}
    \end{minipage}%
    }
    \caption{$\text{MAP@10}$ and $\text{P@10}$ in different sparsity variations of UGEN-V1 Benchmark}
    \label{fig:sparsity-overview}
\end{figure}

One of the benchmark properties that we care about is sparsity, which relates to the amount of null or missing attribute values 
in the tables. Within the \tname framework, this property is controlled by a script that randomly removes values in a table until the desired sparsity is reached. 
\cref{fig:sparsity-overview} illustrates how different methods can handle benchmarks with different rates of sparsity. We vary sparsity from 0\% to 20\% in X-axes and report MAP@10 and P@10 in y-axes. 
SANTOS, Starmie, and Starmie-Vicuna are not impacted by sparsity very much. However, the performance  of the column-based approach (D$^3$L) goes down significantly when we increase sparsity. For example, its MAP@10 reduces by 15 \% when we increase sparsity to 20 \%.  This gives us an interesting insight that the methods which make unionability decisions by capturing the overall table context rather than considering individual columns independently are more tolerant towards sparsity. 

\subsubsection{Topic-Based Analysis}


\begin{table*}
\caption{Top-5 Topics where each method performs the best.}
\begin{tabular}{@{}llllll@{}}
\toprule
    \textbf{Method} &
    \textbf{Top 1} &
    \textbf{Top 2}  & 
    \textbf{Top 3}  & 
    \textbf{Top 4}  & 
    \textbf{Top 5} \\ 
\hline
    \textbf{D$^3$L} &
    Ecology &
    Gardening  & 
    Engineering  & 
    Astronomy  & 
    Computer Science \\ 

    \textbf{SANTOS} &
    Culture &
    Law  & 
    Gardening  & 
    Photography  & 
    Veterinary Medicine \\ 

    \textbf{Starmie} &
    Biology &
    Climatology  & 
    Economics  & 
    Veterinary Medicine  & 
    Photography \\ 

    \textbf{Starmie-Vicuna} &
    Biology &
    Climatology  & 
    Economics  & 
    Photography  & 
    Computer Science \\ 
\bottomrule
\end{tabular}

\label{tab:topic-top5-overview}
\end{table*}

\begin{table*}
\caption{Bottom-5 Topics where each method performs the least.}
\begin{tabular}{@{}llllll@{}}
\toprule
    \textbf{Method} &
    \textbf{Bottom 1} &
    \textbf{Bottom 2}  & 
    \textbf{Bottom 3}  & 
    \textbf{Bottom 4}  & 
    \textbf{Bottom 5} \\ 
\hline
    \textbf{D$^3$L} &
    Psychology &
    Sociology  & 
    Sports  & 
    Veterinary Medicine  & 
    World Geography \\ 

    \textbf{SANTOS} &
    Politics &
    Engineering  & 
    Fashion  & 
    Geology  & 
    World History \\ 

    \textbf{Starmie} &
    Architecture &
    Horticulture  & 
    Environment  & 
    Science  & 
    Religion \\ 

    \textbf{Starmie-Vicuna} &
    Astronomy &
    Geophysics  & 
    Religion  & 
    Business  & 
    Genealogy \\ 
\bottomrule
\end{tabular}
\label{tab:topic-bottom5-overview}
\end{table*}





\begin{figure}[t]
    \centering
\includegraphics[width=.495\textwidth]{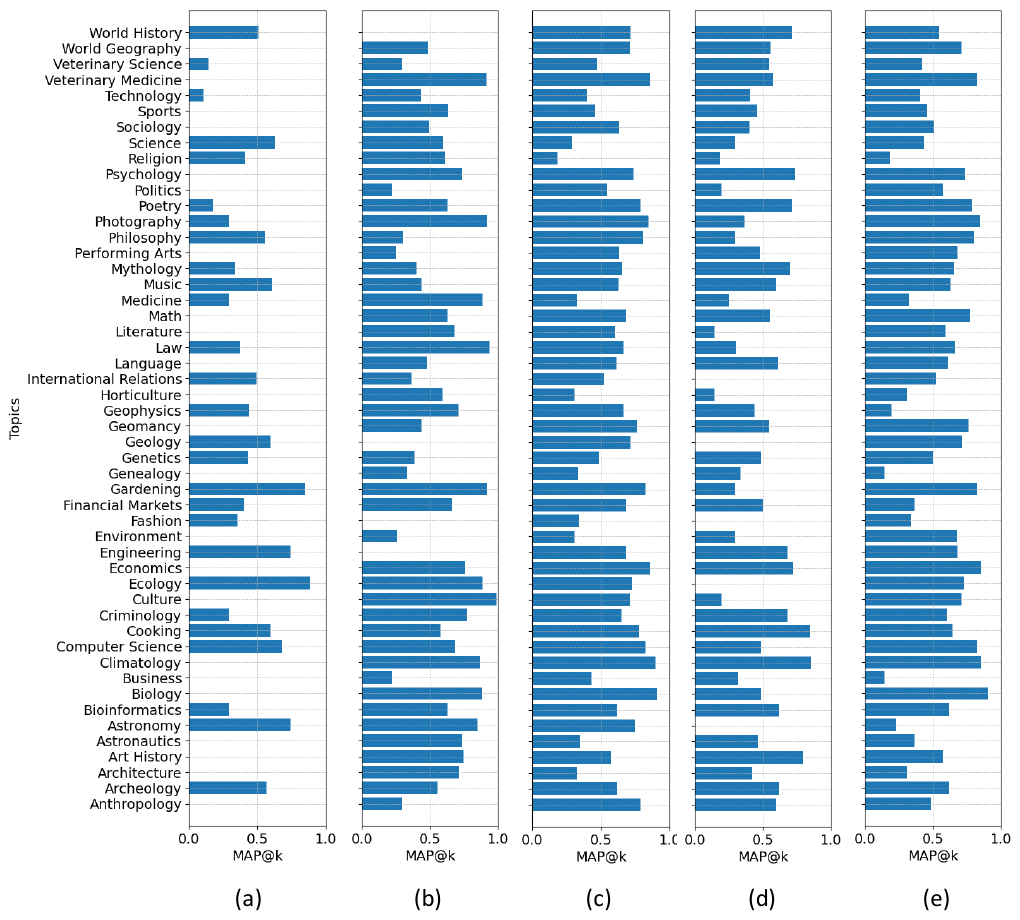}
    \caption{Topic-based MAP@k (k=10) by (a) D$^3$L (b) SANTOS (c) Starmie (d) Starmie-Vicuna$_{Zero}$ (e) Starmie-Vicuna$_{Optim}$ methods in \bname benchmark} 
    \label{fig:topic_specific_all}
\end{figure}

One of the benefits of \tname framework is that we can generate unionable and non-unionable table pairs based on topics. This enables us to perform a topic-based analysis and understand how each baseline performs on various topics. 
In ~\cref{tab:topic-top5-overview} and ~\cref{tab:topic-bottom5-overview}, we show the top-5 and bottom-5 topics out of 50 topics based on MAP@10 for each method. 
For a comprehensive analysis, we also report MAP@10 by each union search method on all 50 topics in ~\cref{fig:topic_specific_all}.

Except for Starmie and its LLM variation (Starmie-Vicuna), it is interesting to see that there is not much overlap in the top-5 best topics of each method. Even more interesting, there is an overlap between the top-5 topics of one method and the bottom-5 topics of another method. For example, on one hand, \texttt{Engineering} is the third best topic for D$^3$L but the second worst topic for \texttt{SANTOS}. On the other hand, D$^3$L has \texttt{Veterinary Medicine} in its bottom-5, while the same topic is among SANTOS' top-5 best topics.
This signifies that each method captures different properties in the table to infer table unionability and an easy topic for one method could be difficult for another method. This result suggests that future work on building domain-specific search methods could be useful along with  combining the strength of multiple methods to build a single highly-effective table union search method. 

Additional interesting insights are visible in \cref{fig:topic_specific_all} which allows a comparison among methods for a given topic. For example, \texttt{World History} achieves a fair MAP@10 score among all methods except SANTOS which achieves a score of 0. This can be attributed to poor coverage of their knowledge base in this topic, which is key for a successful utilization of SANTOS~\cite{2023_khatiwada_santos}. Another apparent example is the topic of \texttt{Climatology} for which D$^3$L fails to find a single unionable table (MAP@10=0) while others were very successful with high MAP@10 scores. Finally, another interesting insight is the topic-specific benefit of in-context learning, comparing  Starmie-Vicuna$ _{Zero}$ (\cref{fig:topic_specific_all} (d)) Starmie-Vicuna$ _{Optim}$ (\cref{fig:topic_specific_all} (e)). For example, along side topics such as \texttt{Bioinformatics} and \texttt{Language} which are unaffected, topics such as \texttt{Gardening} and \texttt{Geology} are significantly improved with ICL and, interestingly, the performance over topics such as \texttt{Cooking} and \texttt{Geophysics} actually decline.    

\subsubsection{Non-Unionable Pair Analysis}\label{sec:nonunionability-analysis}
\begin{figure*}[hbt!]
\includegraphics[width=\textwidth]{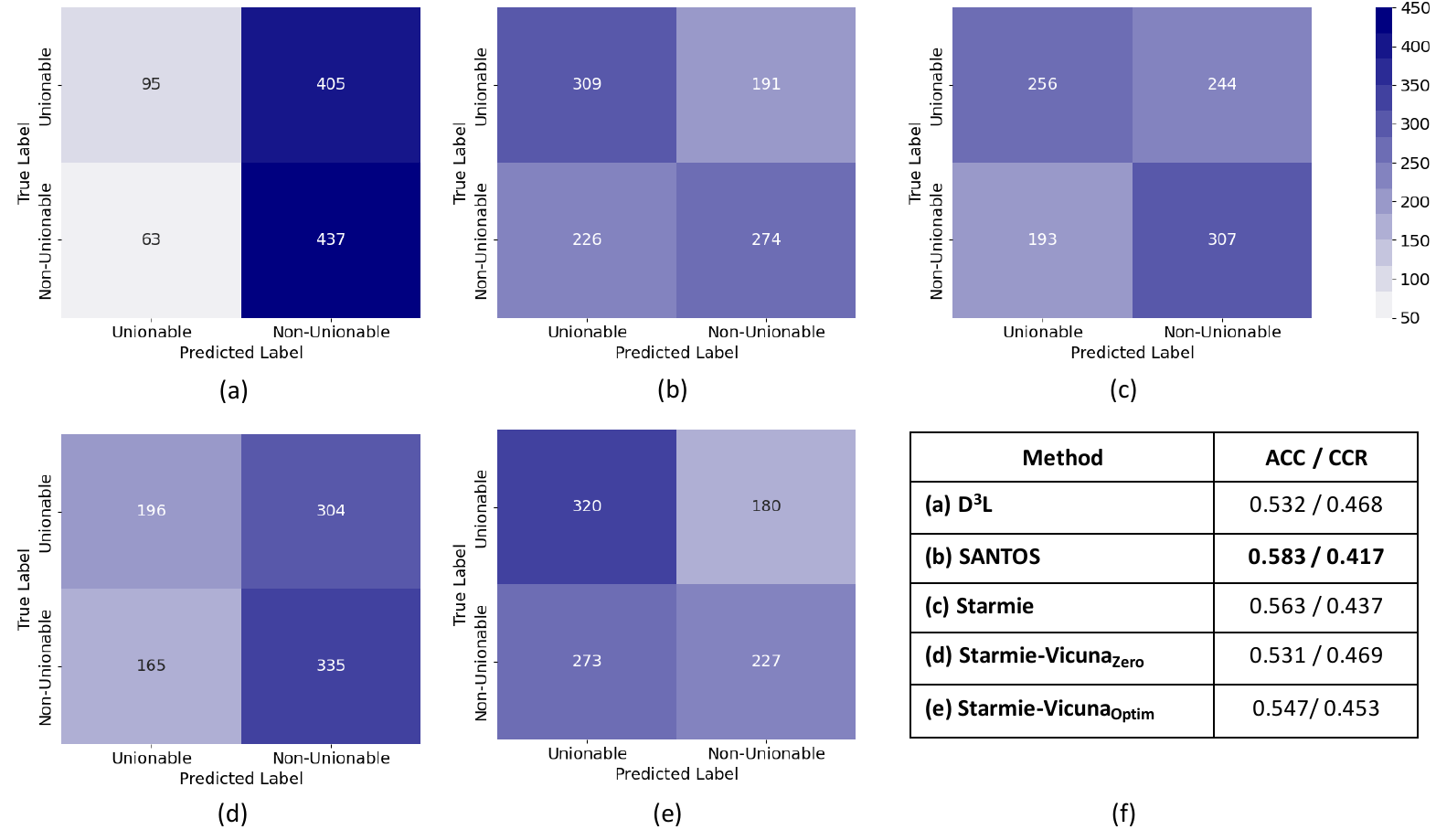}
    \caption{Confusion matrices for (a) D$^3$L (b) SANTOS (c) Starmie (d) Starmie-Vicuna$_{Zero}$ (e) Starmie-Vicuna$_{Optim}$ methods on distinguishing unionable and non-unionable pairs in \bname Benchmark. A legend in the top-right corner is for all the confusion matrices. (f) Accuracy and Corner Case Ratio of different methods in \bname Benchmark}
    \label{fig:confusion_matrix}
\end{figure*}
\tname has the ability to generate labeled non-unionable table pairs from the same topic,
something that was previously overlooked in unionability benchmarks~\cite{nargesian2018table, 2023_khatiwada_santos}.
As reported in~\cref{tab:union-overview}, we have 50 query tables and each query has 10 unionable data lake tables and 10 non-unionable data lake tables, all of which are on the same topic and labeled in the ground truth. 
The other tables that are unlabeled with respect to the query are non-unionable as they are semantically different, not even from the same topic. In this section, we analyze the potential impact of non-unionable table pairs from the same topic on the performance of the union search methods by creating confusion matrices using the 1000 labeled unionable and non-unionable pairs and reporting accuracy over them vs overall non-unionable tables.

To create a confusion matrix, we need true labels, false labels, and predicted labels. True labels and false labels are "unionable" and "non-unionable" table pairs from the ground truth respectively. Note that we use search methods that returns top-k results for each query. As the total number of unionable table pairs for each query is 10 in the ground truth, we search for the top-10 unionable tables for each query. With 50 query tables, we get 500  predicted true labels. The table pairs that are in the ground truth, but do not make it to the top-10 are false (non-unionable) predicted labels. We compare predicted labels with the ground truth to construct confusion matrices.


We report confusion matrix and accuracy in ~\cref{fig:confusion_matrix} for each method in \bname. 
If we see correct non-unionable predictions (bottom-right of confusion matrices), D$^3L$ seems to be the best at detecting non-unionable tables (predicts 437 out of 500 non-unionable pairs accurately).
Notice however, if we look at true positives (top-left of confusion matrix in~\cref{fig:confusion_matrix}(a)), D$^3$L has the least number, around two times 
fewer
than the second-least performing Starmie-Vicuna$_{Zero}$~(\cref{fig:confusion_matrix} (d)).
This means that the false positives in D$^3$L's results are other unlabeled non-unionable tables from random topics rather than the labeled "non-unionable" tables from the same topic. 
But if we look at SANTOS, Starmie, and Starmie's LLM variations that capture the table context to make unionability decisions, they have high true positives and relatively lower true negatives. This means that their false positives are mostly non-unionable tables from the same topic, as we discussed in~\cref{ex:running_example}. 
This is also reflected in the accuracy numbers~\cref{fig:confusion_matrix} (f). For instance, SANTOS seems to be the most balanced method in differentiating unionable and non-unionable among the same topic, achieving the highest accuracy. This is because SANTOS captures fine-grained table context by looking at the binary relationship between the column pairs and getting less confused on the tables from the same topic. On the other hand, Starmie and its LLM variations capture the overall table context in the form of column embeddings without necessarily looking at the binary relationships between the columns explicitly. This may have deprived them of separating the non-unionable from unionable tables within the same topic.
Hence, we get a very interesting insight that table unionability is not just about finding the tables from the same topics. Instead, the search methods should also separate non-unionability among the same topic tables, an observation which has been overlooked in the current benchmarking literature.
\subsection{Discussion and Future Work}

In \tname, we use the massive knowledge present in LLMs, namely GPT3, to generate table pairs that are either unionable or non-unionable for specified topics. However, there are certain limitations posed by this framework. Due to the dependence on LLMs to generate the data, it is not guaranteed that the size and type of table pairs would be exactly as 
requested in a prompt. For instance, it is possible 
that an LLM generates
a table with a different number of rows than requested in the prompt. Furthermore, due to the token limit size, in the case of the Starmie-LLM variations, we rely on a single table row each to identify unionability. 

In~\Cref{sec:new_unionability_benchmark}, we reported that the generation of \bname took about 10 hours and \$18 USD. 
The total runtimes of existing methods on \bname for all 50 queries were about 7 minutes for D$^3$L, 3 minutes for SANTOS and 2 minutes for Starmie.
The Starmie-LLM variation methods had varying query-time based on their model sizes, length of the prompt, and platform for running them. In zero-shot case, Starmie-GPT-2XL took about 0.20 seconds per query (one of 50 queries), Starmie-Vicuna took about 0.53 seconds per query, and Starmie-Alpaca took around 0.54 seconds per query. GPT2-XL is a 1.5 Billion parameter model while Vicuna and Alpaca models are around 7 Billion parameter models. Hence, GPT2-XL had a relatively faster inference time than Vicuna and Alpaca. Furthermore, since Vicuna and Alpaca were similarly sized, they have similar query-times. These query times increase  non-linearly.
Between zero-shot and one-shot, there is a 56\% increase in average query time. Between two-shot and one-shot, there is about 39\% increase in average query time. Lastly, between three-shot and two-shot, there is about 34\% increase in query time. This is consistent for both Vicuna and Alpaca. For GPT2-XL, these rate of increase  are about 19\%, 25\%, and 23\% respectively. While GPT3 is the largest model amongst these models (175 Billion parameter), Starmie-GPT3 took an average of 0.3 seconds per query regardless of the size of the prompt. However, the cost of each prompt inference increased by $\$0.005 \cdot ICL$ for each query where $ICL$ refers to the number of In-Context learning examples provided in the prompt.

\begin{figure}[hbt!]
    \centering
    \includegraphics[width=\linewidth]{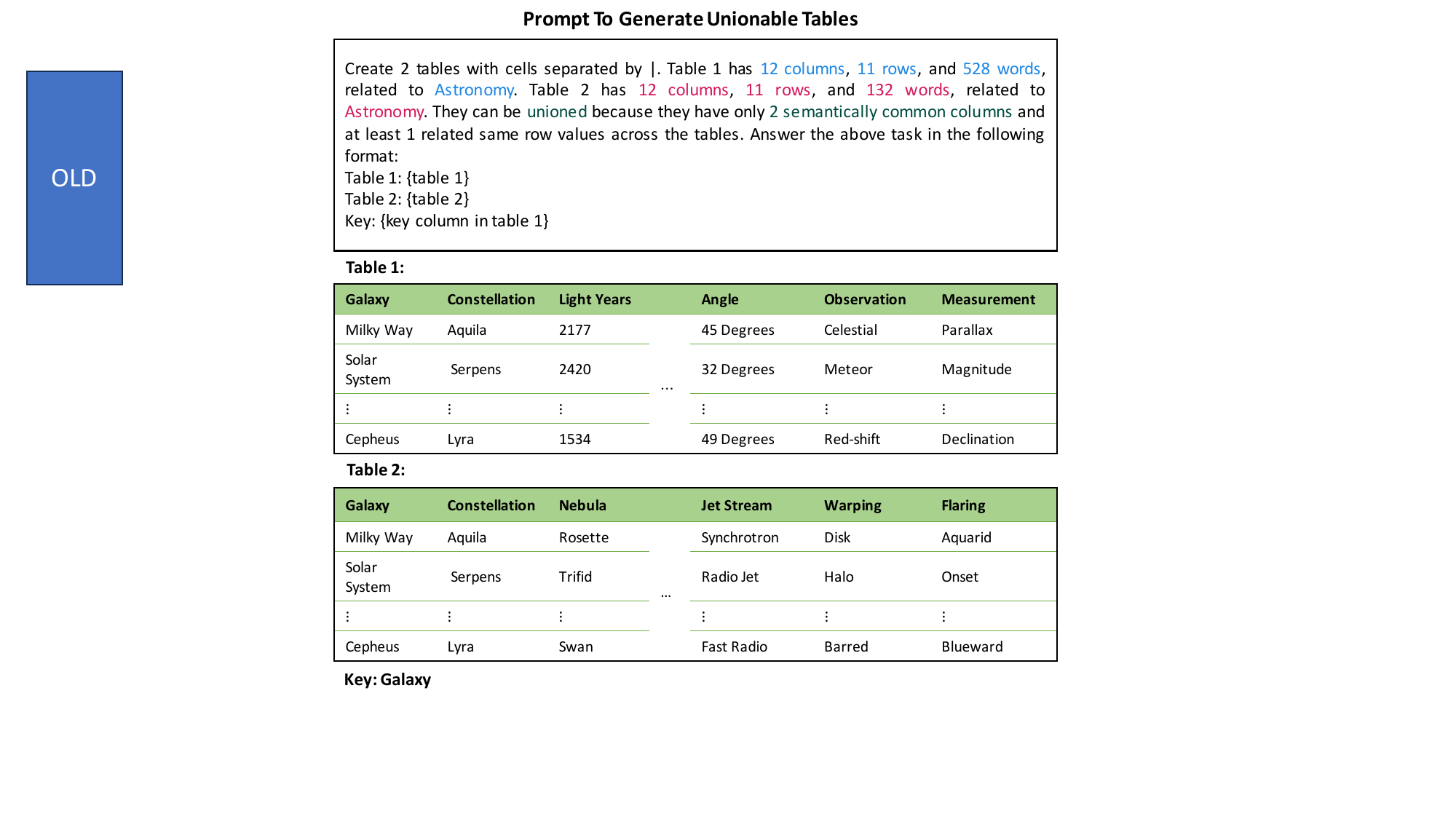}
    \caption{Unionable and Joinable example in \bname}
    \label{fig:joinable_example}
\end{figure}

In \cref{fig:generation-example}, we illustrated a relatively challenging example generated by \tname framework. While such examples do exist in \bname, making it a relatively challenging benchmark, there are also table-pair instances that are relatively easier or are more related to other kinds of inter-table tasks. In \cref{fig:joinable_example}, we illustrate a unionable instance that has a large overlap between the first two columns of each table. Since the overlap includes exact column names, row values, and their relative positions, this example is not a challenging one. Furthermore, this is more related to a joinable table task than a unionable one. Since the prompts in \tname are zero-shot in nature, LLM interprets the unionability task to various degrees. As future work, including examples for each table-pair generation could be more helpful to create more unionability-specific tasks. While this can be limited due by token size limits in LLM prompting, there are many new LLMs such as GPT4~\cite{openai2023gpt4}, that have very recently been released and have higher token limit tolerance. Apart from this direction, the example showcases the ability for LLM to generate other inter-table tasks such as joinability. Hence, \tname framework can be used to generate other inter-table task benchmarks for tasks such as entity-matching and joinability. Lastly, using LLMs as new table union search method relies on current methods such as  Starmie to provide a small set of candidate unionable tables and then classify whether an LLM understands these are unionable or not. As future work, we would like to have a framework where an LLM can be a standalone method to perform table-union search.



\section{Conclusion}
\label{sec:conclusion}
We present \tname, a framework that uses Large Language Models to automatically generate tables for the table union search task. Using \tname, we generate the \bname benchmark, which we showed to be a realistic but more challenging benchmark than existing benchmarks for state-of-the-art table union search methods. It also allows for a more in-depth analysis of these methods, since it generates labeled non-unionable and allows the user to control the topics used.  Our new methodology enabled the first topic-based analysis of table union search and the first in-depth analysis of all methods true/false positive/negative rates. Finally, we have made both \tname and \bname publicly available for other researchers and practitioners to utilize in their experiments and analysis. 
\begin{acks}
This work was supported in part by NSF
under award numbers IIS-1956096 and IIS-2107248.
\end{acks}

\newpage
\bibliographystyle{ACM-Reference-Format}
\bibliography{proposal}


\begin{thebibliography}{45}


\ifx \showCODEN    \undefined \def \showCODEN     #1{\unskip}     \fi
\ifx \showDOI      \undefined \def \showDOI       #1{#1}\fi
\ifx \showISBNx    \undefined \def \showISBNx     #1{\unskip}     \fi
\ifx \showISBNxiii \undefined \def \showISBNxiii  #1{\unskip}     \fi
\ifx \showISSN     \undefined \def \showISSN      #1{\unskip}     \fi
\ifx \showLCCN     \undefined \def \showLCCN      #1{\unskip}     \fi
\ifx \shownote     \undefined \def \shownote      #1{#1}          \fi
\ifx \showarticletitle \undefined \def \showarticletitle #1{#1}   \fi
\ifx \showURL      \undefined \def \showURL       {\relax}        \fi
\providecommand\bibfield[2]{#2}
\providecommand\bibinfo[2]{#2}
\providecommand\natexlab[1]{#1}
\providecommand\showeprint[2][]{arXiv:#2}

\bibitem[\protect\citeauthoryear{??}{com}{[n.d.]}]%
        {commoncrawl}
 \bibinfo{year}{[n.d.]}\natexlab{}.
\newblock \bibinfo{title}{Common Crawl}.
\newblock \bibinfo{howpublished}{https://commoncrawl.org/}.
\newblock


\bibitem[\protect\citeauthoryear{Arocena, Glavic, Ciucanu, and Miller}{Arocena
  et~al\mbox{.}}{2015a}]%
        {2015_arocena_ibench}
\bibfield{author}{\bibinfo{person}{Patricia~C. Arocena}, \bibinfo{person}{Boris
  Glavic}, \bibinfo{person}{Radu Ciucanu}, {and}
  \bibinfo{person}{Ren{\'{e}}e~J. Miller}.} \bibinfo{year}{2015}\natexlab{a}.
\newblock \showarticletitle{The iBench Integration Metadata Generator}.
\newblock \bibinfo{journal}{\emph{Proc. {VLDB} Endow.}} \bibinfo{volume}{9},
  \bibinfo{number}{3} (\bibinfo{year}{2015}), \bibinfo{pages}{108--119}.
\newblock
\urldef\tempurl%
\url{https://doi.org/10.14778/2850583.2850586}
\showDOI{\tempurl}


\bibitem[\protect\citeauthoryear{Arocena, Glavic, Mecca, Miller, Papotti, and
  Santoro}{Arocena et~al\mbox{.}}{2015b}]%
        {2015_arocena_bart}
\bibfield{author}{\bibinfo{person}{Patricia~C. Arocena}, \bibinfo{person}{Boris
  Glavic}, \bibinfo{person}{Giansalvatore Mecca},
  \bibinfo{person}{Ren{\'{e}}e~J. Miller}, \bibinfo{person}{Paolo Papotti},
  {and} \bibinfo{person}{Donatello Santoro}.} \bibinfo{year}{2015}\natexlab{b}.
\newblock \showarticletitle{Messing Up with {BART:} Error Generation for
  Evaluating Data-Cleaning Algorithms}.
\newblock \bibinfo{journal}{\emph{Proc. {VLDB} Endow.}} \bibinfo{volume}{9},
  \bibinfo{number}{2} (\bibinfo{year}{2015}), \bibinfo{pages}{36--47}.
\newblock
\urldef\tempurl%
\url{https://doi.org/10.14778/2850578.2850579}
\showDOI{\tempurl}


\bibitem[\protect\citeauthoryear{Arora, Yang, Eyuboglu, Narayan, Hojel,
  Trummer, and Ré}{Arora et~al\mbox{.}}{2023}]%
        {arora2023language}
\bibfield{author}{\bibinfo{person}{Simran Arora}, \bibinfo{person}{Brandon
  Yang}, \bibinfo{person}{Sabri Eyuboglu}, \bibinfo{person}{Avanika Narayan},
  \bibinfo{person}{Andrew Hojel}, \bibinfo{person}{Immanuel Trummer}, {and}
  \bibinfo{person}{Christopher Ré}.} \bibinfo{year}{2023}\natexlab{}.
\newblock \bibinfo{title}{Language Models Enable Simple Systems for Generating
  Structured Views of Heterogeneous Data Lakes}.
\newblock
\newblock
\showeprint[arxiv]{2304.09433}~[cs.CL]


\bibitem[\protect\citeauthoryear{Bogatu, Fernandes, Paton, and
  Konstantinou}{Bogatu et~al\mbox{.}}{2020}]%
        {DBLP:conf/icde/BogatuFP020}
\bibfield{author}{\bibinfo{person}{Alex Bogatu}, \bibinfo{person}{Alvaro A.~A.
  Fernandes}, \bibinfo{person}{Norman~W. Paton}, {and}
  \bibinfo{person}{Nikolaos Konstantinou}.} \bibinfo{year}{2020}\natexlab{}.
\newblock \showarticletitle{Dataset Discovery in Data Lakes}. In
  \bibinfo{booktitle}{\emph{ICDE}}. \bibinfo{pages}{709--720}.
\newblock


\bibitem[\protect\citeauthoryear{Brinkmann, Primpeli, and Bizer}{Brinkmann
  et~al\mbox{.}}{2023a}]%
        {brinkmann2023web}
\bibfield{author}{\bibinfo{person}{Alexander Brinkmann}, \bibinfo{person}{Anna
  Primpeli}, {and} \bibinfo{person}{Christian Bizer}.}
  \bibinfo{year}{2023}\natexlab{a}.
\newblock \showarticletitle{The Web Data Commons Schema. org Data Set Series}.
  In \bibinfo{booktitle}{\emph{Companion Proceedings of the ACM Web Conference
  2023}}. \bibinfo{pages}{136--139}.
\newblock


\bibitem[\protect\citeauthoryear{Brinkmann, Shraga, Der, and Bizer}{Brinkmann
  et~al\mbox{.}}{2023b}]%
        {brinkmann2023product}
\bibfield{author}{\bibinfo{person}{Alexander Brinkmann}, \bibinfo{person}{Roee
  Shraga}, \bibinfo{person}{Reng~Chiz Der}, {and} \bibinfo{person}{Christian
  Bizer}.} \bibinfo{year}{2023}\natexlab{b}.
\newblock \showarticletitle{Product Information Extraction using ChatGPT}.
\newblock \bibinfo{journal}{\emph{arXiv preprint arXiv:2306.14921}}
  (\bibinfo{year}{2023}).
\newblock


\bibitem[\protect\citeauthoryear{Brown, Mann, Ryder, Subbiah, Kaplan, Dhariwal,
  Neelakantan, Shyam, Sastry, Askell, et~al\mbox{.}}{Brown
  et~al\mbox{.}}{2020}]%
        {brown2020language}
\bibfield{author}{\bibinfo{person}{Tom Brown}, \bibinfo{person}{Benjamin Mann},
  \bibinfo{person}{Nick Ryder}, \bibinfo{person}{Melanie Subbiah},
  \bibinfo{person}{Jared~D Kaplan}, \bibinfo{person}{Prafulla Dhariwal},
  \bibinfo{person}{Arvind Neelakantan}, \bibinfo{person}{Pranav Shyam},
  \bibinfo{person}{Girish Sastry}, \bibinfo{person}{Amanda Askell},
  {et~al\mbox{.}}} \bibinfo{year}{2020}\natexlab{}.
\newblock \showarticletitle{Language models are few-shot learners}.
\newblock \bibinfo{journal}{\emph{Advances in neural information processing
  systems}}  \bibinfo{volume}{33} (\bibinfo{year}{2020}),
  \bibinfo{pages}{1877--1901}.
\newblock


\bibitem[\protect\citeauthoryear{Carlini, Ippolito, Jagielski, Lee, Tramer, and
  Zhang}{Carlini et~al\mbox{.}}{2022}]%
        {carlini2022quantifying}
\bibfield{author}{\bibinfo{person}{Nicholas Carlini}, \bibinfo{person}{Daphne
  Ippolito}, \bibinfo{person}{Matthew Jagielski}, \bibinfo{person}{Katherine
  Lee}, \bibinfo{person}{Florian Tramer}, {and} \bibinfo{person}{Chiyuan
  Zhang}.} \bibinfo{year}{2022}\natexlab{}.
\newblock \showarticletitle{Quantifying Memorization Across Neural Language
  Models}. In \bibinfo{booktitle}{\emph{The Eleventh International Conference
  on Learning Representations}}.
\newblock


\bibitem[\protect\citeauthoryear{Carlini, Tramer, Wallace, Jagielski,
  Herbert-Voss, Lee, Roberts, Brown, Song, Erlingsson, et~al\mbox{.}}{Carlini
  et~al\mbox{.}}{2021}]%
        {carlini2021extracting}
\bibfield{author}{\bibinfo{person}{Nicholas Carlini}, \bibinfo{person}{Florian
  Tramer}, \bibinfo{person}{Eric Wallace}, \bibinfo{person}{Matthew Jagielski},
  \bibinfo{person}{Ariel Herbert-Voss}, \bibinfo{person}{Katherine Lee},
  \bibinfo{person}{Adam Roberts}, \bibinfo{person}{Tom Brown},
  \bibinfo{person}{Dawn Song}, \bibinfo{person}{Ulfar Erlingsson},
  {et~al\mbox{.}}} \bibinfo{year}{2021}\natexlab{}.
\newblock \showarticletitle{Extracting training data from large language
  models}. In \bibinfo{booktitle}{\emph{30th USENIX Security Symposium (USENIX
  Security 21)}}. \bibinfo{pages}{2633--2650}.
\newblock


\bibitem[\protect\citeauthoryear{Chintagunta, Katariya, Amatriain, and
  Kannan}{Chintagunta et~al\mbox{.}}{2021}]%
        {chintagunta2021medically}
\bibfield{author}{\bibinfo{person}{Bharath Chintagunta}, \bibinfo{person}{Namit
  Katariya}, \bibinfo{person}{Xavier Amatriain}, {and} \bibinfo{person}{Anitha
  Kannan}.} \bibinfo{year}{2021}\natexlab{}.
\newblock \showarticletitle{Medically aware GPT-3 as a data generator for
  medical dialogue summarization}. In \bibinfo{booktitle}{\emph{Machine
  Learning for Healthcare Conference}}. PMLR, \bibinfo{pages}{354--372}.
\newblock


\bibitem[\protect\citeauthoryear{Cohen, Geva, Berant, and Globerson}{Cohen
  et~al\mbox{.}}{2023}]%
        {cohen-etal-2023-crawling}
\bibfield{author}{\bibinfo{person}{Roi Cohen}, \bibinfo{person}{Mor Geva},
  \bibinfo{person}{Jonathan Berant}, {and} \bibinfo{person}{Amir Globerson}.}
  \bibinfo{year}{2023}\natexlab{}.
\newblock \showarticletitle{Crawling The Internal Knowledge-Base of Language
  Models}. In \bibinfo{booktitle}{\emph{Findings of the Association for
  Computational Linguistics: EACL 2023}}. \bibinfo{publisher}{Association for
  Computational Linguistics}, \bibinfo{address}{Dubrovnik, Croatia},
  \bibinfo{pages}{1856--1869}.
\newblock
\urldef\tempurl%
\url{https://aclanthology.org/2023.findings-eacl.139}
\showURL{%
\tempurl}


\bibitem[\protect\citeauthoryear{Cutrona, Chen, Efthymiou, Hassanzadeh,
  Jim{\'{e}}nez{-}Ruiz, Sequeda, Srinivas, Abdelmageed, Hulsebos, Oliveira, and
  Pesquita}{Cutrona et~al\mbox{.}}{2021}]%
        {2021_cutrona_semtab}
\bibfield{author}{\bibinfo{person}{Vincenzo Cutrona}, \bibinfo{person}{Jiaoyan
  Chen}, \bibinfo{person}{Vasilis Efthymiou}, \bibinfo{person}{Oktie
  Hassanzadeh}, \bibinfo{person}{Ernesto Jim{\'{e}}nez{-}Ruiz},
  \bibinfo{person}{Juan Sequeda}, \bibinfo{person}{Kavitha Srinivas},
  \bibinfo{person}{Nora Abdelmageed}, \bibinfo{person}{Madelon Hulsebos},
  \bibinfo{person}{Daniela Oliveira}, {and} \bibinfo{person}{Catia Pesquita}.}
  \bibinfo{year}{2021}\natexlab{}.
\newblock \showarticletitle{Results of SemTab 2021}. In
  \bibinfo{booktitle}{\emph{Semantic Web Challenge on Tabular Data to Knowledge
  Graph Matching}}, Vol.~\bibinfo{volume}{3103}. \bibinfo{pages}{1--12}.
\newblock


\bibitem[\protect\citeauthoryear{Dehghani, Zamani, Severyn, Kamps, and
  Croft}{Dehghani et~al\mbox{.}}{2017}]%
        {dehghani2017neural}
\bibfield{author}{\bibinfo{person}{Mostafa Dehghani}, \bibinfo{person}{Hamed
  Zamani}, \bibinfo{person}{Aliaksei Severyn}, \bibinfo{person}{Jaap Kamps},
  {and} \bibinfo{person}{W~Bruce Croft}.} \bibinfo{year}{2017}\natexlab{}.
\newblock \showarticletitle{Neural ranking models with weak supervision}. In
  \bibinfo{booktitle}{\emph{Proceedings of the 40th international ACM SIGIR
  conference on research and development in information retrieval}}.
  \bibinfo{pages}{65--74}.
\newblock


\bibitem[\protect\citeauthoryear{Dong, Li, Dai, Zheng, Wu, Chang, Sun, Xu, and
  Sui}{Dong et~al\mbox{.}}{2022}]%
        {dong2022survey}
\bibfield{author}{\bibinfo{person}{Qingxiu Dong}, \bibinfo{person}{Lei Li},
  \bibinfo{person}{Damai Dai}, \bibinfo{person}{Ce Zheng},
  \bibinfo{person}{Zhiyong Wu}, \bibinfo{person}{Baobao Chang},
  \bibinfo{person}{Xu Sun}, \bibinfo{person}{Jingjing Xu}, {and}
  \bibinfo{person}{Zhifang Sui}.} \bibinfo{year}{2022}\natexlab{}.
\newblock \showarticletitle{A Survey for In-context Learning}.
\newblock \bibinfo{journal}{\emph{arXiv preprint arXiv:2301.00234}}
  (\bibinfo{year}{2022}).
\newblock


\bibitem[\protect\citeauthoryear{Fan, Wang, Li, Zhang, and Miller}{Fan
  et~al\mbox{.}}{2023}]%
        {DBLP:journals/pvldb/FanWLZM23}
\bibfield{author}{\bibinfo{person}{Grace Fan}, \bibinfo{person}{Jin Wang},
  \bibinfo{person}{Yuliang Li}, \bibinfo{person}{Dan Zhang}, {and}
  \bibinfo{person}{Ren{\'{e}}e~J. Miller}.} \bibinfo{year}{2023}\natexlab{}.
\newblock \showarticletitle{Semantics-aware Dataset Discovery from Data Lakes
  with Contextualized Column-based Representation Learning}.
\newblock \bibinfo{journal}{\emph{{PVDLB}}} \bibinfo{volume}{16},
  \bibinfo{number}{7} (\bibinfo{year}{2023}), \bibinfo{pages}{1726--1739}.
\newblock


\bibitem[\protect\citeauthoryear{Gray}{Gray}{1993}]%
        {DBLP:books/mk/Gray93}
\bibfield{editor}{\bibinfo{person}{Jim Gray}} (Ed.).
  \bibinfo{year}{1993}\natexlab{}.
\newblock \bibinfo{booktitle}{\emph{The Benchmark Handbook for Database and
  Transaction Systems (2nd Edition)}}.
\newblock \bibinfo{publisher}{Morgan Kaufmann}.
\newblock
\showISBNx{1-55860-292-5}


\bibitem[\protect\citeauthoryear{Guha, Brickley, and Macbeth}{Guha
  et~al\mbox{.}}{2016}]%
        {guha2016schema}
\bibfield{author}{\bibinfo{person}{Ramanathan~V Guha}, \bibinfo{person}{Dan
  Brickley}, {and} \bibinfo{person}{Steve Macbeth}.}
  \bibinfo{year}{2016}\natexlab{}.
\newblock \showarticletitle{Schema. org: evolution of structured data on the
  web}.
\newblock \bibinfo{journal}{\emph{Commun. ACM}} \bibinfo{volume}{59},
  \bibinfo{number}{2} (\bibinfo{year}{2016}), \bibinfo{pages}{44--51}.
\newblock


\bibitem[\protect\citeauthoryear{Helali, Vashisth, Carrier, Hose, and
  Mansour}{Helali et~al\mbox{.}}{2023}]%
        {2023_helali_linked_data}
\bibfield{author}{\bibinfo{person}{Mossad Helali}, \bibinfo{person}{Shubham
  Vashisth}, \bibinfo{person}{Philippe Carrier}, \bibinfo{person}{Katja Hose},
  {and} \bibinfo{person}{Essam Mansour}.} \bibinfo{year}{2023}\natexlab{}.
\newblock \showarticletitle{Linked Data Science Powered by Knowledge Graphs}.
\newblock \bibinfo{journal}{\emph{CoRR}}  \bibinfo{volume}{abs/2303.02204}
  (\bibinfo{year}{2023}).
\newblock


\bibitem[\protect\citeauthoryear{Hu, Wang, Qin, Lei, Shen, Faloutsos,
  Katsifodimos, Karypis, Wen, and Yu}{Hu et~al\mbox{.}}{2023}]%
        {2023_hu_autotus}
\bibfield{author}{\bibinfo{person}{Xuming Hu}, \bibinfo{person}{Shen Wang},
  \bibinfo{person}{Xiao Qin}, \bibinfo{person}{Chuan Lei},
  \bibinfo{person}{Zhengyuan Shen}, \bibinfo{person}{Christos Faloutsos},
  \bibinfo{person}{Asterios Katsifodimos}, \bibinfo{person}{George Karypis},
  \bibinfo{person}{Lijie Wen}, {and} \bibinfo{person}{Philip~S. Yu}.}
  \bibinfo{year}{2023}\natexlab{}.
\newblock \showarticletitle{Automatic Table Union Search with Tabular
  Representation Learning}. In \bibinfo{booktitle}{\emph{Findings of the
  Association for Computational Linguistics: {ACL} 2023, Toronto, Canada, July
  9-14, 2023}}. \bibinfo{publisher}{Association for Computational Linguistics},
  \bibinfo{pages}{3786--3800}.
\newblock
\urldef\tempurl%
\url{https://aclanthology.org/2023.findings-acl.233}
\showURL{%
\tempurl}


\bibitem[\protect\citeauthoryear{Hulsebos, Demiralp, and Groth}{Hulsebos
  et~al\mbox{.}}{2023}]%
        {2023_hulsebos_gittables}
\bibfield{author}{\bibinfo{person}{Madelon Hulsebos},
  \bibinfo{person}{{\c{C}}agatay Demiralp}, {and} \bibinfo{person}{Paul
  Groth}.} \bibinfo{year}{2023}\natexlab{}.
\newblock \showarticletitle{GitTables: {A} Large-Scale Corpus of Relational
  Tables}.
\newblock \bibinfo{journal}{\emph{Proc. {ACM} Manag. Data}}
  \bibinfo{volume}{1}, \bibinfo{number}{1} (\bibinfo{year}{2023}),
  \bibinfo{pages}{30:1--30:17}.
\newblock
\urldef\tempurl%
\url{https://doi.org/10.1145/3588710}
\showDOI{\tempurl}


\bibitem[\protect\citeauthoryear{Khatiwada, Fan, Shraga, Chen, Gatterbauer,
  Miller, and Riedewald}{Khatiwada et~al\mbox{.}}{2023}]%
        {2023_khatiwada_santos}
\bibfield{author}{\bibinfo{person}{Aamod Khatiwada}, \bibinfo{person}{Grace
  Fan}, \bibinfo{person}{Roee Shraga}, \bibinfo{person}{Zixuan Chen},
  \bibinfo{person}{Wolfgang Gatterbauer}, \bibinfo{person}{Renée~J. Miller},
  {and} \bibinfo{person}{Mirek Riedewald}.} \bibinfo{year}{2023}\natexlab{}.
\newblock \showarticletitle{SANTOS: Relationship-based Semantic Table Union
  Search}. In \bibinfo{booktitle}{\emph{Accepted to appear in SIGMOD
  Conference}}. \bibinfo{publisher}{ACM}.
\newblock
\urldef\tempurl%
\url{https://arxiv.org/pdf/2209.13589.pdf}
\showURL{%
\tempurl}


\bibitem[\protect\citeauthoryear{Koutras, Siachamis, Ionescu, Psarakis, Brons,
  Fragkoulis, Lofi, Bonifati, and Katsifodimos}{Koutras et~al\mbox{.}}{2021}]%
        {DBLP:conf/icde/KoutrasSIPBFLBK21}
\bibfield{author}{\bibinfo{person}{Christos Koutras}, \bibinfo{person}{George
  Siachamis}, \bibinfo{person}{Andra Ionescu}, \bibinfo{person}{Kyriakos
  Psarakis}, \bibinfo{person}{Jerry Brons}, \bibinfo{person}{Marios
  Fragkoulis}, \bibinfo{person}{Christoph Lofi}, \bibinfo{person}{Angela
  Bonifati}, {and} \bibinfo{person}{Asterios Katsifodimos}.}
  \bibinfo{year}{2021}\natexlab{}.
\newblock \showarticletitle{Valentine: Evaluating Matching Techniques for
  Dataset Discovery}. In \bibinfo{booktitle}{\emph{{ICDE}}}.
  \bibinfo{pages}{468--479}.
\newblock


\bibitem[\protect\citeauthoryear{Leventidis, Rocco, Gatterbauer, Miller, and
  Riedewald}{Leventidis et~al\mbox{.}}{2021}]%
        {DBLP:conf/edbt/LeventidisRMRG21}
\bibfield{author}{\bibinfo{person}{Aristotelis Leventidis},
  \bibinfo{person}{Laura~Di Rocco}, \bibinfo{person}{Wolfgang Gatterbauer},
  \bibinfo{person}{Ren{\'{e}}e~J. Miller}, {and} \bibinfo{person}{Mirek
  Riedewald}.} \bibinfo{year}{2021}\natexlab{}.
\newblock \showarticletitle{DomainNet: Homograph Detection for Data Lake
  Disambiguation}. In \bibinfo{booktitle}{\emph{{EDBT}}},
  \bibfield{editor}{\bibinfo{person}{Yannis Velegrakis},
  \bibinfo{person}{Demetris Zeinalipour{-}Yazti}, \bibinfo{person}{Panos~K.
  Chrysanthis}, {and} \bibinfo{person}{Francesco Guerra}} (Eds.).
  \bibinfo{pages}{13--24}.
\newblock


\bibitem[\protect\citeauthoryear{Liu, Shen, Zhang, Dolan, Carin, and Chen}{Liu
  et~al\mbox{.}}{2021}]%
        {liu2021makes}
\bibfield{author}{\bibinfo{person}{Jiachang Liu}, \bibinfo{person}{Dinghan
  Shen}, \bibinfo{person}{Yizhe Zhang}, \bibinfo{person}{Bill Dolan},
  \bibinfo{person}{Lawrence Carin}, {and} \bibinfo{person}{Weizhu Chen}.}
  \bibinfo{year}{2021}\natexlab{}.
\newblock \bibinfo{title}{What Makes Good In-Context Examples for GPT-$3$?}
\newblock
\newblock
\showeprint[arxiv]{2101.06804}~[cs.CL]


\bibitem[\protect\citeauthoryear{Liu, Ott, Goyal, Du, Joshi, Chen, Levy, Lewis,
  Zettlemoyer, and Stoyanov}{Liu et~al\mbox{.}}{2019}]%
        {liu2019roberta}
\bibfield{author}{\bibinfo{person}{Yinhan Liu}, \bibinfo{person}{Myle Ott},
  \bibinfo{person}{Naman Goyal}, \bibinfo{person}{Jingfei Du},
  \bibinfo{person}{Mandar Joshi}, \bibinfo{person}{Danqi Chen},
  \bibinfo{person}{Omer Levy}, \bibinfo{person}{Mike Lewis},
  \bibinfo{person}{Luke Zettlemoyer}, {and} \bibinfo{person}{Veselin
  Stoyanov}.} \bibinfo{year}{2019}\natexlab{}.
\newblock \showarticletitle{Roberta: A robustly optimized bert pretraining
  approach}.
\newblock \bibinfo{journal}{\emph{arXiv preprint arXiv:1907.11692}}
  (\bibinfo{year}{2019}).
\newblock


\bibitem[\protect\citeauthoryear{Nargesian, Zhu, Pu, and Miller}{Nargesian
  et~al\mbox{.}}{2018}]%
        {nargesian2018table}
\bibfield{author}{\bibinfo{person}{Fatemeh Nargesian}, \bibinfo{person}{Erkang
  Zhu}, \bibinfo{person}{Ken~Q Pu}, {and} \bibinfo{person}{Ren{\'e}e~J
  Miller}.} \bibinfo{year}{2018}\natexlab{}.
\newblock \showarticletitle{Table union search on open data}.
\newblock \bibinfo{journal}{\emph{PVLDB}} \bibinfo{volume}{11},
  \bibinfo{number}{7} (\bibinfo{year}{2018}), \bibinfo{pages}{813--825}.
\newblock


\bibitem[\protect\citeauthoryear{OpenAI}{OpenAI}{[n.d.]}]%
        {openaichat}
\bibfield{author}{\bibinfo{person}{OpenAI}.} \bibinfo{year}{[n.d.]}\natexlab{}.
\newblock \bibinfo{title}{Chat GPT}.
\newblock \bibinfo{howpublished}{https://chat.openai.com/chat}.
\newblock


\bibitem[\protect\citeauthoryear{OpenAI}{OpenAI}{2023}]%
        {openai2023gpt4}
\bibfield{author}{\bibinfo{person}{OpenAI}.} \bibinfo{year}{2023}\natexlab{}.
\newblock \bibinfo{title}{GPT-4 Technical Report}.
\newblock
\newblock
\showeprint[arxiv]{2303.08774}~[cs.CL]


\bibitem[\protect\citeauthoryear{Patterson}{Patterson}{2012}]%
        {patterson2012technical}
\bibfield{author}{\bibinfo{person}{David Patterson}.}
  \bibinfo{year}{2012}\natexlab{}.
\newblock \showarticletitle{Technical perspective for better or worse,
  benchmarks shape a field}.
\newblock \bibinfo{journal}{\emph{Commun. ACM}} \bibinfo{volume}{55},
  \bibinfo{number}{7} (\bibinfo{year}{2012}).
\newblock


\bibitem[\protect\citeauthoryear{Peeters and Bizer}{Peeters and Bizer}{2023}]%
        {peeters2023using}
\bibfield{author}{\bibinfo{person}{Ralph Peeters} {and}
  \bibinfo{person}{Christian Bizer}.} \bibinfo{year}{2023}\natexlab{}.
\newblock \showarticletitle{Using ChatGPT for Entity Matching}.
\newblock \bibinfo{journal}{\emph{arXiv preprint arXiv:2305.03423}}
  (\bibinfo{year}{2023}).
\newblock


\bibitem[\protect\citeauthoryear{Primpeli and Bizer}{Primpeli and
  Bizer}{2020}]%
        {primpeli2020profiling}
\bibfield{author}{\bibinfo{person}{Anna Primpeli} {and}
  \bibinfo{person}{Christian Bizer}.} \bibinfo{year}{2020}\natexlab{}.
\newblock \showarticletitle{Profiling entity matching benchmark tasks}. In
  \bibinfo{booktitle}{\emph{CIKM}}. \bibinfo{pages}{3101--3108}.
\newblock


\bibitem[\protect\citeauthoryear{Shraga, Roitman, Feigenblat, and
  Cannim}{Shraga et~al\mbox{.}}{2020}]%
        {shraga2020web}
\bibfield{author}{\bibinfo{person}{Roee Shraga}, \bibinfo{person}{Haggai
  Roitman}, \bibinfo{person}{Guy Feigenblat}, {and} \bibinfo{person}{Mustafa
  Cannim}.} \bibinfo{year}{2020}\natexlab{}.
\newblock \showarticletitle{Web table retrieval using multimodal deep
  learning}. In \bibinfo{booktitle}{\emph{Proceedings of the 43rd international
  ACM SIGIR conference on research and development in information retrieval}}.
  \bibinfo{pages}{1399--1408}.
\newblock


\bibitem[\protect\citeauthoryear{Srinivas, Dolby, Abdelaziz, Hassanzadeh,
  Kokel, Khatiwada, Pedapati, Chaudhury, and Samulowitz}{Srinivas
  et~al\mbox{.}}{2023}]%
        {srinivas2023lakebench}
\bibfield{author}{\bibinfo{person}{Kavitha Srinivas}, \bibinfo{person}{Julian
  Dolby}, \bibinfo{person}{Ibrahim Abdelaziz}, \bibinfo{person}{Oktie
  Hassanzadeh}, \bibinfo{person}{Harsha Kokel}, \bibinfo{person}{Aamod
  Khatiwada}, \bibinfo{person}{Tejaswini Pedapati}, \bibinfo{person}{Subhajit
  Chaudhury}, {and} \bibinfo{person}{Horst Samulowitz}.}
  \bibinfo{year}{2023}\natexlab{}.
\newblock \showarticletitle{LakeBench: Benchmarks for Data Discovery over Data
  Lakes}.
\newblock \bibinfo{journal}{\emph{arXiv preprint arXiv:2307.04217}}
  (\bibinfo{year}{2023}).
\newblock


\bibitem[\protect\citeauthoryear{Taori, Gulrajani, Zhang, Dubois, Li, Guestrin,
  Liang, and Hashimoto}{Taori et~al\mbox{.}}{2023}]%
        {alpaca}
\bibfield{author}{\bibinfo{person}{Rohan Taori}, \bibinfo{person}{Ishaan
  Gulrajani}, \bibinfo{person}{Tianyi Zhang}, \bibinfo{person}{Yann Dubois},
  \bibinfo{person}{Xuechen Li}, \bibinfo{person}{Carlos Guestrin},
  \bibinfo{person}{Percy Liang}, {and} \bibinfo{person}{Tatsunori~B.
  Hashimoto}.} \bibinfo{year}{2023}\natexlab{}.
\newblock \bibinfo{title}{Stanford Alpaca: An Instruction-following LLaMA
  model}.
\newblock
  \bibinfo{howpublished}{\url{https://github.com/tatsu-lab/stanford_alpaca}}.
\newblock


\bibitem[\protect\citeauthoryear{Trummer}{Trummer}{2022a}]%
        {trummer2022codexdb}
\bibfield{author}{\bibinfo{person}{Immanuel Trummer}.}
  \bibinfo{year}{2022}\natexlab{a}.
\newblock \showarticletitle{CodexDB: Synthesizing code for query processing
  from natural language instructions using GPT-3 Codex}.
\newblock \bibinfo{journal}{\emph{Proceedings of the VLDB Endowment}}
  \bibinfo{volume}{15}, \bibinfo{number}{11} (\bibinfo{year}{2022}),
  \bibinfo{pages}{2921--2928}.
\newblock


\bibitem[\protect\citeauthoryear{Trummer}{Trummer}{2022b}]%
        {trummer2022bert}
\bibfield{author}{\bibinfo{person}{Immanuel Trummer}.}
  \bibinfo{year}{2022}\natexlab{b}.
\newblock \showarticletitle{From BERT to GPT-3 codex: harnessing the potential
  of very large language models for data management}.
\newblock \bibinfo{journal}{\emph{Proceedings of the VLDB Endowment}}
  \bibinfo{volume}{15}, \bibinfo{number}{12} (\bibinfo{year}{2022}),
  \bibinfo{pages}{3770--3773}.
\newblock


\bibitem[\protect\citeauthoryear{Veyseh, Nguyen, Min, and Nguyen}{Veyseh
  et~al\mbox{.}}{2021}]%
        {Veyseh2021AugmentingOE}
\bibfield{author}{\bibinfo{person}{Amir Pouran~Ben Veyseh},
  \bibinfo{person}{Minh~Le Nguyen}, \bibinfo{person}{Bonan Min}, {and}
  \bibinfo{person}{Thien~Huu Nguyen}.} \bibinfo{year}{2021}\natexlab{}.
\newblock \showarticletitle{Augmenting Open-Domain Event Detection with
  Synthetic Data from GPT-2}. In \bibinfo{booktitle}{\emph{ECML/PKDD}}.
\newblock


\bibitem[\protect\citeauthoryear{{WDC}}{{WDC}}{2017}]%
        {WDC_t2d}
{WDC} \bibinfo{year}{2017}\natexlab{}.
\newblock \bibinfo{title}{{T2D Gold Standard}}.
\newblock
\newblock
\newblock
\shownote{http://webdatacommons.org/webtables/goldstandard.html.}


\bibitem[\protect\citeauthoryear{West, Bhagavatula, Hessel, Hwang, Jiang,
  Le~Bras, Lu, Welleck, and Choi}{West et~al\mbox{.}}{2022}]%
        {west-etal-2022-symbolic}
\bibfield{author}{\bibinfo{person}{Peter West}, \bibinfo{person}{Chandra
  Bhagavatula}, \bibinfo{person}{Jack Hessel}, \bibinfo{person}{Jena Hwang},
  \bibinfo{person}{Liwei Jiang}, \bibinfo{person}{Ronan Le~Bras},
  \bibinfo{person}{Ximing Lu}, \bibinfo{person}{Sean Welleck}, {and}
  \bibinfo{person}{Yejin Choi}.} \bibinfo{year}{2022}\natexlab{}.
\newblock \showarticletitle{Symbolic Knowledge Distillation: from General
  Language Models to Commonsense Models}. In
  \bibinfo{booktitle}{\emph{Proceedings of the 2022 Conference of the North
  American Chapter of the Association for Computational Linguistics: Human
  Language Technologies}}. \bibinfo{publisher}{Association for Computational
  Linguistics}, \bibinfo{address}{Seattle, United States},
  \bibinfo{pages}{4602--4625}.
\newblock
\urldef\tempurl%
\url{https://doi.org/10.18653/v1/2022.naacl-main.341}
\showDOI{\tempurl}


\bibitem[\protect\citeauthoryear{Wolf, Debut, Sanh, Chaumond, Delangue, Moi,
  Cistac, Rault, Louf, Funtowicz, et~al\mbox{.}}{Wolf et~al\mbox{.}}{2020}]%
        {wolf2020transformers}
\bibfield{author}{\bibinfo{person}{Thomas Wolf}, \bibinfo{person}{Lysandre
  Debut}, \bibinfo{person}{Victor Sanh}, \bibinfo{person}{Julien Chaumond},
  \bibinfo{person}{Clement Delangue}, \bibinfo{person}{Anthony Moi},
  \bibinfo{person}{Pierric Cistac}, \bibinfo{person}{Tim Rault},
  \bibinfo{person}{R{\'e}mi Louf}, \bibinfo{person}{Morgan Funtowicz},
  {et~al\mbox{.}}} \bibinfo{year}{2020}\natexlab{}.
\newblock \showarticletitle{Transformers: State-of-the-art natural language
  processing}. In \bibinfo{booktitle}{\emph{Proceedings of the 2020 conference
  on empirical methods in natural language processing: system demonstrations}}.
  \bibinfo{pages}{38--45}.
\newblock


\bibitem[\protect\citeauthoryear{Xian, Lampert, Schiele, and Akata}{Xian
  et~al\mbox{.}}{2018}]%
        {xian2018zero}
\bibfield{author}{\bibinfo{person}{Yongqin Xian}, \bibinfo{person}{Christoph~H
  Lampert}, \bibinfo{person}{Bernt Schiele}, {and} \bibinfo{person}{Zeynep
  Akata}.} \bibinfo{year}{2018}\natexlab{}.
\newblock \showarticletitle{Zero-shot learning—a comprehensive evaluation of
  the good, the bad and the ugly}.
\newblock \bibinfo{journal}{\emph{IEEE transactions on pattern analysis and
  machine intelligence}} \bibinfo{volume}{41}, \bibinfo{number}{9}
  (\bibinfo{year}{2018}), \bibinfo{pages}{2251--2265}.
\newblock


\bibitem[\protect\citeauthoryear{Yang, Malaviya, Fernandez, Swayamdipta, Bras,
  Wang, Bhagavatula, Choi, and Downey}{Yang et~al\mbox{.}}{2020}]%
        {yang2020generative}
\bibfield{author}{\bibinfo{person}{Yiben Yang}, \bibinfo{person}{Chaitanya
  Malaviya}, \bibinfo{person}{Jared Fernandez}, \bibinfo{person}{Swabha
  Swayamdipta}, \bibinfo{person}{Ronan~Le Bras}, \bibinfo{person}{Ji-Ping
  Wang}, \bibinfo{person}{Chandra Bhagavatula}, \bibinfo{person}{Yejin Choi},
  {and} \bibinfo{person}{Doug Downey}.} \bibinfo{year}{2020}\natexlab{}.
\newblock \showarticletitle{Generative data augmentation for commonsense
  reasoning}.
\newblock \bibinfo{journal}{\emph{arXiv preprint arXiv:2004.11546}}
  (\bibinfo{year}{2020}).
\newblock


\bibitem[\protect\citeauthoryear{Zhao, Wallace, Feng, Klein, and Singh}{Zhao
  et~al\mbox{.}}{2021}]%
        {zhao2021calibrate}
\bibfield{author}{\bibinfo{person}{Tony~Z. Zhao}, \bibinfo{person}{Eric
  Wallace}, \bibinfo{person}{Shi Feng}, \bibinfo{person}{Dan Klein}, {and}
  \bibinfo{person}{Sameer Singh}.} \bibinfo{year}{2021}\natexlab{}.
\newblock \bibinfo{title}{Calibrate Before Use: Improving Few-Shot Performance
  of Language Models}.
\newblock
\newblock
\showeprint[arxiv]{2102.09690}~[cs.CL]


\bibitem[\protect\citeauthoryear{Zheng, Chiang, Sheng, Zhuang, Wu, Zhuang, Lin,
  Li, Li, Xing, Zhang, Gonzalez, and Stoica}{Zheng et~al\mbox{.}}{2023}]%
        {vicuna}
\bibfield{author}{\bibinfo{person}{Lianmin Zheng}, \bibinfo{person}{Wei-Lin
  Chiang}, \bibinfo{person}{Ying Sheng}, \bibinfo{person}{Siyuan Zhuang},
  \bibinfo{person}{Zhanghao Wu}, \bibinfo{person}{Yonghao Zhuang},
  \bibinfo{person}{Zi Lin}, \bibinfo{person}{Zhuohan Li},
  \bibinfo{person}{Dacheng Li}, \bibinfo{person}{Eric.~P Xing},
  \bibinfo{person}{Hao Zhang}, \bibinfo{person}{Joseph~E. Gonzalez}, {and}
  \bibinfo{person}{Ion Stoica}.} \bibinfo{year}{2023}\natexlab{}.
\newblock \bibinfo{title}{Judging LLM-as-a-judge with MT-Bench and Chatbot
  Arena}.
\newblock
\newblock
\showeprint[arxiv]{2306.05685}~[cs.CL]


\end{thebibliography}
\end{document}